\documentclass[prb,twocolumn,showpacs,amsmath,amssymb]{revtex4}

\usepackage{graphicx}
\usepackage{dcolumn}
\usepackage{bm}
\usepackage{amsmath}

\begin{document}
\title
{{\em Ab Initio} Calculations of $\bm H_{c2}$ in Type-II Superconductors:\\
Basic Formalism and Model Calculations}

\author{Takafumi Kita}
\affiliation{Division of Physics, Hokkaido University, Sapporo 060-0810, Japan}

\author{Masao Arai}
\affiliation{
National Institute for Materials Science, Namiki 1-1, Tsukuba, Ibaraki 305-0044, Japan}

\date{\today}

\begin{abstract}
Detailed Fermi-surface structures are essential 
to describe the upper critical field $H_{c2}$ in type-II superconductors, 
as first noticed by Hohenberg and Werthamer 
[Phys.\ Rev.\ {\bf 153},\ 493 (1967)] and shown explicitly by
Butler for high-purity cubic Niobium [Phys.\ Rev.\ Lett.\ {\bf 44},\ 1516 (1980)].
We derive an $H_{c2}$ equation for classic type-II superconductors
which is applicable to systems with anisotropic Fermi surfaces and/or energy gaps
under arbitrary field directions.
It can be solved efficiently by using Fermi surfaces
from {\em ab initio} electronic-structure calculations.
Thus, it is expected to enhance our quantitative understanding on $H_{c2}$.
Based on the formalism, we calculate $H_{c2}$ curves for
Fermi surfaces of a three-dimensional tight-binding model
with cubic symmetry, an isotropic gap, and no impurity scatterings.
It is found that, as the Fermi surface approaches to the Brillouin zone
boundary, the reduced critical field $h^{*}(T/T_{c})$,
which is normalized by the initial slope at $T_{c}$,
is enhanced significantly over the curve for the spherical Fermi surface
with a marked upward curvature. Thus, the Fermi-surface anisotropy can
be a main source of the upward curvature in $H_{c2}$ near $T_c$.
\end{abstract}
\pacs{74.25.Op, 71.18.+y}
\maketitle

\section{Introduction}
\label{sec:intro}

The upper critical field $H_{c2}$ is one of the most fundamental quantities 
in type-II superconductors.
After the pioneering work by Abrikosov\cite{Abrikosov57}
based on the Ginzburg-Landau (GL) equations,\cite{GL}
theoretical efforts have been made for its quantitative description at all 
temperatures.\cite{Gor'kov59-2,Maki64,deGennes64,HW66,WHH66,Maki66,
HW67,Eilenberger67,EA67,WM67,GG68,
Mattheiss70,TN70,Takanaka71,RB73,Takanaka75,Teichler75,Pohl76,EP76,Butler80-1,Butler80-2,
YK80,SK80,SS84,SS86,PS87,
Prohammer90,RS90,RSS91,Tesanovic91,Langmann92,PS93,NMA95,Kita98-2,Maniv01,YK02,Miranovic03,Dahm03,Kita03}
However, we still have a limited success
when compared with those for the electronic structures 
in the normal state.\cite{Kohn99}
The purpose of the present paper is to provide a theoretical framework
which enables us {\em ab initio} calculations of $H_{c2}$ as 
accurate as electronic-structure calculations in the normal state.

Necessary ingredients to be included are
(i) nonlocal effects effective at low temperatures;
(ii) impurity scattering;
(iii) Fermi-surface anisotropy;
(iv) strong-coupling effects;
(v) gap anisotropy;
(vi) mixing of higher Landau levels in the
spatial dependence of the pair potential;
(vii) Landau-level quantization in the quasiparticle energy;\cite{GG68,Tesanovic91,
NMA95,Kita98-2,Maniv01,YK02}
(viii) fluctuations beyond the mean-field theory.\cite{Blatter94}
We here derive an $H_{c2}$ equation
which is numerically tractable, including all the effects except (vii) and (viii).

An $H_{c2}$ equation considering the effects (i) and (ii) was
obtained by Helfand and Werthamer.\cite{HW66} 
It was extended by Hohenberg and Werthamer\cite{HW67}
to take the Fermi-surface anisotropy (iii) into account.
Equations with the strong-coupling effects (iv) were derived
by Eilenberger and Ambegaokar\cite{EA67} using Matsubara frequencies
and by Werthamer and McMillan\cite{WM67} on the real energy axis,
which are equivalent to one another.
Schossmann and Schachinger\cite{SS86} later incorporated Pauli paramagnetism
into the strong-coupling equation.
Although an equation including (i)-(iv) was presented by Langmann,\cite{Langmann92}
it is still rather complicated for carrying out an actual numerical computation.
On the other hand, Rieck and Scharnberg\cite{RS90} presented 
an efficient $H_{c2}$ equation 
where the effects (i)-(iii) and (vi) were taken into account,
and also (v) in the special case of the clean limit.
See also the work by Rieck, Scharnberg, and Schopohl\cite{RSS91}
where the strong-coupling effects (v) have also been considered.
Our study can be regarded as a direct extension of 
the Rieck-Scharnberg equation\cite{RS90} 
to incorporate (i)-(iv) simultaneously.
To this end, we adopt a slightly different and (probably) more convenient
procedure of using creation and annihilation operators.
We will proceed with clarifying the connections with 
the Rieck-Scharnberg equation as explicitly as possible.

The remarkable success of the simplified Bardeen-Cooper-Schrieffer (BCS) theory\cite{BCS,Parks69}
tells us that detailed electronic structures are rather irrelevant to the
properties of classic superconductors at $H\!=\!0$.
However, this is not the case for the properties of type-II superconductors
in finite magnetic fields, especially in the clean limit,
as first recognized by Hohenberg and Werthamer.\cite{HW67}
Their effort to include the Fermi-surface anisotropy in 
the $H_{c2}$ equation was motivated by the fact that
the Helfand-Werthamer theory\cite{HW66} 
using the spherical Fermi surface
shows neither qualitative nor
quantitative agreements
with experiments on clean type-II superconductors like Nb\cite{Tilley64,MS65,FSS66} 
and V.\cite{RK66}
Indeed, angular variation in $H_{c2}$ by $10\%$ was
observed at low temperatures in high-quality Nb\cite{Tilley64,Williamson70,Sauerzopf87}
and V\cite{Williamson70,Sauerzopf87} with cubic symmetry.\cite{Weber77}
Also, the reduced critical field
\begin{equation}
h^{*}(t) \equiv  \frac{H_{c2}(t)}{-dH_{c2}(t)/dt|_{t=1}} 
\hspace{5mm} (t \equiv  T/T_{c})\, ,
\label{H^*}
\end{equation}
calculated by Helfand and Werthamer\cite{HW66}
has $h^{*}(0)\!=\!0.727$ in the clean limit,
whereas a later experiment on high-purity Nb shows 
$\langle h^{*}(0)\rangle \!=\!1.06$ for the average over field directions.\cite{Sauerzopf87}
Hohenberg and Werthamer\cite{HW67} carried out 
a perturbation expansion for cubic materials
with respect to the nonlocal correction
where the Fermi-surface anisotropy enters.
They could thereby provide a qualitative understanding of the $H_{c2}$ anisotropy and 
the enhancement of $\langle h^{*}(t)\rangle$ observed in Nb.
They also derived an expression for $\langle h^{*}(0)\rangle$
applicable to anisotropic Fermi surfaces.
It was later used by Mattheiss\cite{Mattheiss70}
to estimate $\langle h^{*}(0)\rangle \!=\! 0.989$ for Nb based on his detailed 
electronic-structure calculation.
The strong dependence of $h^{*}(t)$ in the clean limit
on detailed Fermi-surface structures can also be
seen clearly in the numerical results from a model calculation 
by Rieck and Scharnberg,\cite{RS90}
and from the difference $h^{*}(0)\!=\!0.727$ and $0.591$
between spherical and cylindrical Fermi surfaces, respectively.\cite{Kita03}

\begin{table*}
\caption{\label{tab:table1}Equation numbers for the relevant analytic expressions to
calculate $H_{c2}$. The upper critical field $H_{c2}$ corresponds to the point where
the smallest eigenvalue of the Hermitian matrix ${\cal A}\!=\!({\cal A}_{NN'})$ takes zero.}
\begin{ruledtabular}
\begin{tabular}{ccccccccccccccccccccc}
\vspace{0.5mm}
$\langle \cdots \rangle$ & $\phi({\bf k}_{\rm F})$ & $B$ & $\Phi_{0}$ & $T_{c}$ & $l_{c}$ & $\bar{v}_{{\rm F}+}$   &  $c_{1,2}$ & $\chi_{ij}$
& $\tilde{\varepsilon}_{n}'$ & $\bar{\beta}$ & ${\cal A}_{NN'}$ & ${\cal K}_{NN'}$ & $\eta(x)$ & 
$H_{c2}(T\!\alt\!T_{c})$ & $B_{1}$ & $B_{2}$ & $R$ & $w_{\nu,\mu}$
\\
\hline
(\ref{FSA}) & (\ref{V_kk'}) & (\ref{H-B}) & $hc/2e$ & (\ref{Tc}) & (\ref{l_c})
& (\ref{v_+})  & (\ref{c_12}) & (\ref{chi})  & (\ref{varEpsilon})
& (\ref{beta}) & (\ref{calA}) & (\ref{K_NN'-2}) &(\ref{P_N-BP_N}) &  (\ref{Hc2}) 
& (\ref{B_1}) & (\ref{B_2}) & (\ref{R}) & (\ref{Ws}) &\\
\end{tabular}
\end{ruledtabular}
\end{table*}

On the other hand, it was shown by Werthamer and McMillan\cite{WM67}
that the strong-coupling effects change $h^{*}(t)$ by only $ \alt\! 2\%$
for the spherical Fermi surface
and cannot be the main reason for the enhancement of $h^{*}(0)$ in Nb.

The most complete calculation including the effects (i)-(iv) was 
performed on pure Nb by Butler.\cite{Butler80-1,Butler80-2}
He solved the strong-coupling equation by Eilenberger and
Ambegaokar,\cite{EA67} taking full account of 
the Fermi-surface structure and the phonon spectra 
from his electronic-structure calculations.
He could thereby obtain an excellent agreement with experiments
by Williamson\cite{Williamson70} with $\langle h^{*}(0)\rangle \!=\!0.96$
and by Kerchner {\em et al}.\cite{Kerchner80}
However, a later experiment by Sauerzopf {\em et al}.\ \cite{Sauerzopf87}
on a high-purity Nb shows a larger value $\langle h^{*}(0)\rangle \!=\!1.06$,
thereby suggesting that there may be some factors missing
in Butler's calculation.

Theoretical considerations on the effects (v) and (vi) started much later. 
It was Takanaka\cite{Takanaka75} and 
Teichler\cite{Teichler75,Pohl76} who first included
gap anisotropy (v) in the $H_{c2}$ equation.
They both considered the nonlocal effect perturbatively
adopting a separable pair potential.
Takanaka studied $H_{c2}$ anisotropy observed in uniaxial crystals, 
whereas Teichler applied his theory to the $H_{c2}$ anisotropy in cubic Nb.
This approach by Teichler was extended by Prohammer and Schachinger\cite{PS87} to
anisotropic polycrystals and used by Weber {\em et al}.\ \cite{Weber91}
to analyze anisotropy effects in Nb.

The mixing of higher Landau levels (vi) was first considered by Takanaka
and Nagashima\cite{TN70} in extending the Hohenberg-Werthamer theory
for cubic materials\cite{HW67}
to higher orders in the nonlocal correction.
It was also taken into account by Takanaka\cite{Takanaka75}
in the above-mentioned work,
by Youngner and Klemm\cite{YK80} 
in their perturbation expansion
with respect to the nonlocal corrections,
by Scharnberg and Klemm\cite{SK80} in studying $H_{c2}$ for
$p$-wave superconductors, by Rieck and Scharnberg\cite{RS90}
for superconductors with nearly cylindrical model Fermi surfaces,
and by Prohammer and Carbotte\cite{Prohammer90} for $d$-wave superconductors.
See also a recent work by Miranovi\'c, Machida, and Kogan 
on MgB$_{2}$.\cite{Miranovic03}
Although it plays an important role in the presence 
of gap anisotropy,\cite{SK80,Prohammer90}
this mixing was not considered by Teichler.\cite{Teichler75,Pohl76}

Now, one may be convinced that calculations including (i)-(vi) are still absent.
Especially, many of the theoretical efforts
have been focused only on the special case of cubic materials.\cite{HW67,TN70,Teichler75,
Pohl76,Butler80-1,Butler80-2}
For example, a detailed theory is still absent for the large positive (upward) 
curvature observed in $H_{c2}(T\!\alt\! T_{c})$ of layered 
superconductors,\cite{Muto73,Woollam74}
except a qualitative description by Takanaka\cite{Takanaka75} and 
Dalrymple and Prober.\cite{DP84}
Based on these observations, we here derive an $H_{c2}$ equation
which is numerically tractable for arbitrary crystal structures
and field directions by using Fermi surfaces
from {\em ab initio} electronic-structure calculations.
This kind of calculations has been performed only for Nb by Butler 
so far.\cite{Butler80-1,Butler80-2}
Making such calculations possible for other materials is expected to enhance
our quantitative understanding on $H_{c2}$ substantially.

This paper is organized as follows.
Section II considers the weak-coupling model 
with gap anisotropy and $s$-wave impurity scattering.
We derive an $H_{c2}$ equation valid at all temperatures
as well as an analytic expression for $H_{c2}(T\!\alt\!T_{c})$
up to second order in $1\!-\!T/T_{c}$.
The main analytic results of Sec.\ II are listed in Table I for an easy reference.
Section III extends the $H_{c2}$ equation so as to include 
$p$-wave impurity scattering, spin-orbit impurity scattering,
and strong electron-phonon interactions.
Section IV presents numerical examples for model Fermi surfaces
of a three-dimensional tight-binding model with cubic symmetry.
Section V summarizes the paper.
We put $k_{\rm B}\!=\! 1$ throughout.

\section{Weak-coupling $\bm H_{c2}$ equation}
\subsection{Fermi-surface harmonics and gap anisotropy}

We first specify the gap anisotropy in our consideration
with respect to the Fermi-surface harmonics.
The Fermi-surface harmonics were introduced by Allen\cite{Allen76} as convenient polynomials
in solving the Boltzmann and Eliashberg equations.
They were later used by Langmann\cite{Langmann92} 
to derive an $H_{c2}$ equation applicable to
anisotropic Fermi surfaces and anisotropic pairing interactions.
However, the polynomials constructed by Allen based on the Gram-Schmidt orthonormalization
are not very convenient for treating the gap anisotropy. 
We here adopt an alternative construction starting from the pairing interaction 
$V({\bf k}_{\rm F},{\bf k}_{\rm F}^{\prime})$ on the Fermi surface,\cite{comment1}
where ${\bf k}_{\rm F}$ denotes the Fermi wavevector.
Evidently $V({\bf k}_{\rm F},{\bf k}_{\rm F}^{\prime})$ is Hermitian
$V^{*}({\bf k}_{\rm F},{\bf k}_{\rm F}^{\prime})\!=\!
V({\bf k}_{\rm F}^{\prime},{\bf k}_{\rm F})$,
and invariant under every symmetry operation $R$
of the group $G$ for the relevant crystal as 
$R V({\bf k}_{\rm F},{\bf k}_{\rm F}^{\prime})R^{-1}\!=\!
V({\bf k}_{\rm F},{\bf k}_{\rm F}^{\prime})$.
We hence consider the following eigenvalue problem:
\begin{equation}
\int\!\! dS_{{\rm F}}^{\prime} \rho({\bf k}_{{\rm F}}^{\prime}) \,
V({\bf k}_{\rm F},{\bf k}_{\rm F}^{\prime}) \,
\phi^{(\Gamma j)}_{\gamma}  ({\bf k}_{\rm F}') =
V^{(\Gamma j)}  \phi^{(\Gamma j)}_{\gamma} ({\bf k}_{\rm F}) \, .
\label{EigenFS}
\end{equation}
Here $dS_{{\rm F}}$ denotes an infinitesimal area on the Fermi surface and
$\rho({\bf k}_{{\rm F}})\!\equiv\! [(2\pi)^{3}N(0)|{\bf v}_{{\rm F}}|]^{-1}$
with ${\bf v}_{{\rm F}}$ the Fermi velocity and
$N(0)$ the density of states per one spin and per unit volume at the Fermi energy
in the normal state.
The superscript $\Gamma$ denotes an irreducible representation of $G$,
$j$ distinguishes different eigenvalues belonging to $\Gamma$, and
$\gamma$ specifies an eigenvector in $(\Gamma,j)$.
This eigenvalue problem was also considered by Pokrovskii\cite{Pokrovskii61}
without specifying the symmetry.
The basis functions thereby obtained naturally have all the properties
of Fermi-surface harmonics introduced by Allen.
Especially, they satisfy the orthonormality
and completeness:
\begin{subequations}
\label{completeness}
\begin{equation}
\langle \phi^{(\Gamma j)*}_{\gamma} 
\phi^{(\Gamma' j')}_{\gamma'}  \rangle  = \delta_{\Gamma\Gamma'}
\delta_{jj'}\delta_{\gamma\gamma'} \, ,
\end{equation}
\begin{equation}
\sum_{\Gamma j\gamma} \,
\phi^{(\Gamma j)}_{\gamma}  ({\bf k}_{\rm F})
\phi^{(\Gamma j)*}_{\gamma}  ({\bf k}_{\rm F}') = 
\frac{\delta^{2}({\bf k}_{\rm F}\!-\!{\bf k}_{\rm F}')}{\rho({\bf k}_{{\rm F}}) } \, ,
\end{equation}
\end{subequations}
where $\langle\cdots\rangle$ denotes the Fermi-surface average:
\begin{equation}
\langle A \rangle \equiv
\int\!\! dS_{{\rm F}} \rho({\bf k}_{{\rm F}}) \,
A ({\bf k}_{\rm F}) \, .
\label{FSA}
\end{equation}
Using Eqs.\ (\ref{EigenFS}) and (\ref{completeness}), 
we obtain an alternative expression for the dimensionless pairing interaction 
$\lambda({\bf k}_{\rm F},{\bf k}_{\rm F}')\!\equiv\! 
-N(0)V({\bf k}_{\rm F},{\bf k}_{\rm F}^{\prime})$ as
\begin{equation}
\lambda({\bf k}_{\rm F},{\bf k}_{\rm F}^{\prime})= 
\sum_{\Gamma j \gamma} \lambda^{(\Gamma j)}  \phi^{(\Gamma j)}_{\gamma} ({\bf k}_{\rm F})
\phi^{(\Gamma j)*}_{\gamma} ({\bf k}_{\rm F}') \, .
\label{V_kk'}
\end{equation}
Thus, it is always possible to express a general pairing interaction 
as a sum of separable interactions.
Notice that the above procedure is applicable also to multiband superconductors.
Indeed, we only have to extend the integration over ${\bf k}_{\rm F}$ 
to all the Fermi surfaces.

The Fermi-surface harmonics can be constructed also from the coupling function
$\lambda({\bf k}_{\rm F},{\bf k}_{\rm F}',\varepsilon_{n}\!-\!\varepsilon_{n}')
\!-\!\mu^{*}({\bf k}_{\rm F},{\bf k}_{\rm F}')$
in the strong-coupling Eliashberg theory,\cite{Eliashberg60,Allen82}
where $\varepsilon_{n}\!\equiv\!(2n+1)\pi T$ is the Matsubara energy.
Indeed, we only have to specify an appropriate bosonic Matsubara energy 
$\omega_{l}\!\equiv\! 2l\pi T$ and set $V({\bf k}_{\rm F},{\bf k}_{\rm F}')\!\equiv\!
-[\lambda({\bf k}_{\rm F},{\bf k}_{\rm F}',\omega_{l})
\!-\!\mu^{*}({\bf k}_{\rm F},{\bf k}_{\rm F}')]/N(0)$ in 
Eqs.\ (\ref{EigenFS}) and (\ref{completeness}).
We thereby obtain an alternative expression for the coupling function as
\begin{eqnarray}
&&\hspace{-3mm}
\lambda({\bf k}_{\rm F},{\bf k}_{\rm F}',\varepsilon_{n}\!-\!\varepsilon_{n}')
\!-\!\mu^{*}({\bf k}_{\rm F},{\bf k}_{\rm F}')
\nonumber \\
&&\hspace{-7mm} = 
\sum_{\Gamma j \gamma} [\,\lambda^{(\Gamma j)}(\varepsilon_{n}\!-\!\varepsilon_{n}') 
\!-\! \mu^{*(\Gamma j)}\,]\, \phi^{(\Gamma j)}_{\gamma} ({\bf k}_{\rm F})
\phi^{(\Gamma j)*}_{\gamma} ({\bf k}_{\rm F}') \, .
\end{eqnarray}
We expect that this construction does not depend on the choice of $\omega_{l}$
substantially.
It is worth noting that {\em ab initio} calculations
of the coupling function are now possible 
for phonon-mediated superconductors, as performed recently
for MgB$_{2}$.\cite{Choi02}
Hence {\em ab initio} constructions of the Fermi-surface harmonics by Eq.\ (\ref{EigenFS})
can be carried out in principle.

From now on we consider the cases where (i)
the system has inversion symmetry and (ii) 
a single $\lambda^{(\Gamma j)}$ is relevant which belongs to an even-parity one-dimensional 
representation $\Gamma$.
Indeed, these conditions are met for most superconductors.
Hereafter we will drop all the indices as 
$ \phi^{(\Gamma j)}_{\gamma} ({\bf k}_{\rm F})\!\rightarrow\!  
\phi ({\bf k}_{\rm F})$, for example, and choose 
$\phi({\bf k}_{\rm F})$ as a real function.

\subsection{Eilenberger equations}

Now, let us derive an $H_{c2}$ equation for the second-order transition
in the weak-coupling model 
with $s$-wave impurity scattering
based on the quasiclassical Eilenberger equations.\cite{Eilenberger68,LO68,SR83} 
The Eilenberger equations are derived from the Gor'kov equations 
by assuming a constant density of states near the Fermi energy in the normal state
and integrating out an irrelevant energy variable.\cite{Eilenberger68,LO68,SR83}
Thus, phenomena closely connected with either the energy dependence of 
the density of states\cite{SS84}
or the discreteness in the quasiparticle energy levels\cite{GG68,Tesanovic91,
NMA95,Kita98-2,Maniv01,YK02} are beyond the scope of the present consideration.
We also do not consider Josephson vortices appearing in very anisotropic
layered superconductors.\cite{Koshelev99}
Within the limitations, however, 
the Eilenberger equations provide one of the most convenient starting points
for deriving an $H_{c2}$ equation, as seen below.
This approach was also adopted by Rieck {\em et al}.\cite{RS90,RSS91}

We take the external magnetic field ${\bf H}$ 
along the $z$ axis.
In the presence of Pauli paramagnetism, the average flux density $B$ 
in the bulk is connected with $H$ as
$H\!=\!B\!-\!4\pi \chi_{\rm n} B$, 
where $\chi_{\rm n}$ is the normal-state spin susceptibility.
The fact that $\chi_{\rm n}$ is multiplied by $B$ rather than $H$
corresponds to the fact that the spins respond to the true magnetic field in the bulk.
It hence follows that $B$ is enhanced over $H$ as
\begin{equation}
B=H/(1\!-\!4\pi \chi_{\rm n}) \, .
\label{H-B}
\end{equation}
The vector potential in the bulk at $H\!=\! H_{c2}$ can be written 
accordingly as
\begin{equation}
{\bf A}({\bf r}) =  (0,Bx,0) \, .
\label{A}
\end{equation}
The field ${\bf H}$ is supposed to be along the direction 
$(\sin\theta\cos\varphi,\sin\theta\sin\varphi,\cos\theta)$
in the crystallographic coordinates $(X,Y,Z)$.
The two coordinate systems are connected by the rotation matrix
\begin{equation}
{\cal R}\equiv\left[
\begin{array}{ccc}
\cos\theta\cos\varphi & \cos\theta\sin\varphi & -\sin\theta
\\
-\sin\varphi & \cos\varphi & 0
\\
\sin\theta\cos\varphi & \sin\theta\sin\varphi & \cos\theta
\end{array}\right] ,
\label{calR}
\end{equation}
as ${\cal R}{\bf H}\!=\!(0,0,H)^{\rm T}$, where $^{\rm T}$ denotes transpose. 
We assume that the vortex lattice is uniform along $z$.

With the gap anisotropy specified by $\phi({\bf k}_{{\rm F}})$
and in the presence of Pauli paramagnetism, 
the Eilenberger equations read
\begin{subequations}
\label{Eilens}
\begin{equation}
\left(\!\varepsilon_{n}\!-\! i\mu_{\rm B}B\!+\!\frac{\hbar}{2\tau}\langle g\rangle
\!+\!\frac{1}{2}\hbar{\bf v}_{{\rm F}}\!\cdot\!{\bm \partial}\!\right)\! f=
\left(\!\phi \Delta \!+\!\frac{\hbar}{2\tau}\langle f\rangle\!\right)\! g \, ,
\label{Eilen}
\end{equation}
\begin{equation}
\Delta({\bf r}) \ln \!\frac{T_{c0}}{T}= \pi T \! \sum_{n=-\infty}^{\infty}
\left[\frac{\Delta({\bf r})}{|\varepsilon_{n}|}-\langle 
\phi({\bf k}_{\rm F})f(\varepsilon_{n},{\bf k}_{\rm F},{\bf r})\rangle 
\right]  .
\label{pair}
\end{equation}
\end{subequations}
Here $\mu_{\rm B}$ is the Bohr magneton, 
$\tau$ is the relaxation time by nonmagnetic impurity scattering
in the second-Born approximation,
$\Delta({\bf r})$ is the pair potential,
and ${\bm \partial}$ is defined by
\begin{equation}
{\bm \partial}\equiv {\bm \nabla}-i\frac{2\pi}{\Phi_{0}}{\bf A} \, ,
\end{equation}
with $\Phi_{0}\!\equiv\! hc/2e$ the flux quantum.
We will consider positively charged particles following the convention;
the case of electrons can be obtained directly 
by ${\bf A}\!\rightarrow\!-{\bf A}$, i.e., reversing the magnetic-field direction.
The quasiclassical Green's functions $f$ and $g$ are connected by
$g\!=\!(1\!-\! ff^{\dagger})^{1/2}{\rm sgn}(\varepsilon_{n})$
with $f^{\dagger}(\varepsilon_{n},{\bf k}_{\rm F},{\bf r})\!=
\!f^{*}(-\varepsilon_{n},{\bf k}_{\rm F},{\bf r})$,\cite{Kita03-2}
and $T_{c0}$ denotes the transition temperature in the clean limit $\tau\!=\!\infty$.

To obtain $B_{c2}$, we formally expand
the quasiclassical Green's functions up to the first order in $\Delta$ as
$f\!=\!f^{(1)}$ and $g\!=\! {\rm sgn}(\varepsilon_{n})$.
Substituting the expressions into Eqs.\ (\ref{Eilen}) and (\ref{pair}),
we obtain the linearized self-consistency equations as
\begin{subequations}
\label{Eilens(1)}
\begin{equation}
\left[\,\tilde{\varepsilon}_{n}'
+\frac{{\rm sgn}(\varepsilon_{n})}{2}\hbar{\bf v}_{{\rm F}}\!\cdot\!{\bm \partial}\,\right] 
f^{(1)}=\phi \Delta +\frac{\hbar}{2\tau}\langle f^{(1)}\rangle \, ,
\label{Eilen(1)}
\end{equation}
\begin{equation}
\Delta \ln \!\frac{T_{c0}}{T}= -\pi T \sum_{n=-\infty}^{\infty}
\left(\langle 
\phi f^{(1)}\rangle 
-\frac{\Delta}{|\varepsilon_{n}|}\right)  ,
\label{pair(1)}
\end{equation}
\end{subequations}
with
\begin{equation}
\tilde{\varepsilon}_{n}'\equiv 
\tilde{\varepsilon}_{n}-i\mu_{\rm B}B{\rm sgn}(\varepsilon_{n})\, ,
\hspace{5mm}
\tilde{\varepsilon}_{n}\equiv |\varepsilon_{n}|+\frac{\hbar}{2\tau} \, .
\label{varEpsilon}
\end{equation}

\subsection{Operators and basis functions}

It is useful to transform the gradient operator in Eq.\ (\ref{Eilen}) as
\begin{equation}
{\bf v}_{{\rm F}}\!\cdot\!{\bm \partial}
= ({\bar{v}_{{\rm F}+}^{*} a - \bar{v}_{{\rm F}+} a^{\dagger}})/{\sqrt{2}l_{\rm c}} \, .
\label{gradient}
\end{equation}
Here $l_{\rm c}$ denotes $\frac{1}{\sqrt{2}}$ times the magnetic length
as
\begin{equation}
l_{\rm c}\equiv \sqrt{\Phi_{0}/2\pi B}=\sqrt{\hbar c/2eB} \, .
\label{l_c}
\end{equation}
The operators $a$ and $a^{\dagger}$ are defined by
\begin{equation}
\left[
	\begin{array}{c}
	a \\ a^{\dagger}
	\end{array}
\right]
=\frac{l_{{\rm c}}}{\sqrt{2}}
\left[
	\begin{array}{cc}
	c_{1} & ic_{2} \\
	-c_{1}^{*} & ic_{2}^{*}
	\end{array}
\right]
\left[
	\begin{array}{c}
	\partial_{x} \\ \partial_{y}
	\end{array}
\right] \, ,
\label{aa*}
\end{equation}
where the constants $c_{1}$ and $c_{2}$ are constrained by
\begin{equation}
c_{1}c_{2}^{*}+c_{1}^{*}c_{2}= 2 \, ,
\label{c_12-const}
\end{equation}
so that $[a,a^{\dagger}]\!=\! 1$.
Finally, $\bar{v}_{{\rm F}+}$ is defined by
\begin{equation}
\label{v_+}
\bar{v}_{{\rm F}+}\equiv c_{2}v_{{\rm F}x}+i c_{1}v_{{\rm F}y} \, .
\end{equation}
The constants ($c_{1}$,$c_{2}$) can be fixed conveniently by requiring that
the gradient term 
in the Ginzburg-Landau equation be expressed in terms of $a^{\dagger}a$
without using $aa$ and $a^{\dagger}a^{\dagger}$, i.e.,
the pair potential near $T_{c}$ be described in terms of the lowest Landau level only.
As shown in Appendix A, this condition yields
\begin{subequations}
\label{c_12}
\begin{equation}
c_{1}=\left(\!\frac{\chi_{xx}^{2}}{\chi_{xx} \chi_{yy} -\chi_{xy}^{2}}\!\right)^{\!\!\! 1/4}
\label{c_1}
\end{equation}
\begin{equation}
c_{2}=\left(\!\frac{\chi_{yy}^{2}}{\chi_{xx} \chi_{yy} -\chi_{xy}^{2}}\!\right)^{\!\!\! 1/4}
\exp\!\!\left(\! i\tan^{-1}\!\frac{
-\chi_{xy}}{\sqrt{
\chi_{xx}\chi_{yy}
-\chi_{xy}^{2}}}\!\right) ,
\label{c_2}
\end{equation}
\end{subequations}
where $\chi_{ij}\!\equiv\!\chi_{ij}(T_{c})$ is defined by
\begin{eqnarray}
&&\hspace{-3mm} \chi_{ij}\equiv 
\frac{24(\pi T_{c})^{3}}{7\zeta(3) \langle v_{{\rm F}}^{2}\rangle}
\sum_{n=0}^{\infty} \frac{1}{\tilde{\varepsilon}_{n}^{3}}
\biggl[\,\langle \phi^{2}v_{{\rm F}i}\, v_{{\rm F}j}\rangle
+\frac{\langle \phi\rangle\langle \phi\, v_{{\rm F}i}\, v_{{\rm F}j}\rangle}
{2\tau\varepsilon_{n}} 
\nonumber \\
&&\hspace{6mm} 
+\frac{\langle \phi\rangle\langle \phi v_{{\rm F}i}\, v_{{\rm F}j}\rangle}
{2\tau\varepsilon_{n}} 
+\frac{\langle \phi\rangle\langle \phi\rangle
\langle v_{{\rm F}i}\, v_{{\rm F}j}\rangle}{(2\tau\varepsilon_{n})^{2}} 
\,\biggr]\, ,
\label{chi}
\end{eqnarray}
with $\zeta$ the Riemann zeta function.
Notice that $\chi_{ij}$ is dimensionless,
approaching to $\delta_{ij}$ as $\tau\!\rightarrow\!\infty$ 
for the spherical Fermi surface.
It is a direct generalization of the $\chi$ function
introduced by Gor'kov\cite{Gor'kov59} to anisotropic systems.

The operators in Eq.\ (\ref{aa*}) extends 
$(a_{-},a_{+})$ introduced by Helfand and Werthamer\cite{HW66}
for anisotropic crystals.
For uniaxial crystals, they reduce to the operators used by Takanaka.\cite{Takanaka75}

Using Eq.\ (\ref{aa*}), we can also make up
a set of basis functions to describe
vortex-lattice structures as\cite{Kita98}
\begin{eqnarray}
&&\hspace{-4mm} \psi_{N{\bf q}}({\bf r})=
\sqrt{\frac{2\pi l_{\rm c}}{c_{1}a_{2}\sqrt{\pi }\,V }}
\sum_{n=-{\cal N}_{{\rm f}}/2+1}^{{\cal N}_{{\rm f}}/2}
\exp\!\left[
i q_{y}\!\left(y+ \frac{l_{\rm c}^{2}q_{x}}{2}\right)\!\right]
\nonumber \\
&&\hspace{11mm}
\times\exp\!\left[
i \frac{na_{1x}}{l_{\rm c}^{2}}\!\left(y+
l_{\rm c}^{2}q_{x}- \frac{na_{1y}}{2}\right)
\!\right]
\nonumber \\
&&\hspace{11mm}
\times \exp\!\left[
-\frac{c_{1}c_{2}}{2}\left(\! 
\frac{x- l_{\rm c}^{2}q_{y}- na_{1x}}{c_{1}l_{\rm c}}\!\right)^{\!\! 2} \right] 
\nonumber \\
&&\hspace{11mm}\times 
\frac{1}{\sqrt{2^N N!}}H_{N}\!\!\left(\! 
\frac{x- l_{\rm c}^{2}q_{y}- na_{1x}}{c_{1}l_{\rm c}}\!\right) \, .
\label{basis}
\end{eqnarray}
Here $N\!=\!0,1,2,\cdots$ denotes the Landau level,
${\bf q}$ is an arbitrary chosen magnetic Bloch vector 
characterizing the broken translational symmetry of the vortex lattice
and specifying the core locations, and
$V$ is the volume of the system.
The quantities
$a_{1x}$ and $a_{2}$ are the components of the basic vectors
${\bf a}_{1}$ and ${\bf a}_{2}$ in the $xy$ plane, respectively,
with ${\bf a}_{2}\!\parallel\!\hat{\bf y}$
and $a_{1x}a_{2}\!=\!2\pi l_{{\rm c}}^{2}$,
${\cal N}_{{\rm f}}^{2}$ denotes the number of the flux quantum 
in the system, and
$H_{N}(x)\!\equiv {\rm e}^{x^{2}}\!\left(-\frac{d}{dx}\!\right)^{\! N}
{\rm e}^{-x^{2}}$ is the Hermite polynomial.
The basis functions are both orthonormal and complete, satisfying
$a\psi_{N{\bf q}}\!=\!\sqrt{N}\psi_{N-1{\bf q}}$
and $a^{\dagger}\psi_{N{\bf q}}\!=\!\sqrt{N\!+\! 1}\psi_{N+1{\bf q}}$.

The function (\ref{basis}) is a direct generalization of 
the Eilenberger function\cite{Eilenberger67} 
$\psi_N({\bf r}|{\bf r}_{0})$
with $c_{1}\!=\!c_{2}\!=\! 1$
to anisotropic Fermi surfaces and energy gaps. 
For ${\bf q}\!=\!0$ in the clean limit,
Eq.\ (\ref{basis}) reduces to the function obtained by Rieck {\em et al.},\cite{RS90,RSS91,comment2}
However, they derived it without recource to 
the creation and annihilation operators of Eq.\ (\ref{aa*}).
These operators have simplified the derivation of the basis functions and will also
make the whole calculations below much easier and transparent.

\subsection{Analytic expression of $\bm H_{c2}$ near $T_{c}$}

Using Eq.\ (\ref{aa*}), it is also possible to obtain an analytic expression for 
$B_{c2}\!\equiv\! H_{c2}/(1\!-\! 4\pi \chi_{\rm n})$ near $T_{c}$.
Let us express it as
\begin{equation}
B_{c2}=B_{1}(1-t)+B_{2}(1-t)^{2} \, ,
\label{Hc2}
\end{equation}
with $t\!\equiv\!T/T_{c}$.
The coefficients $B_{1}$ and $B_{2}$ determine the initial slope
and the curvature, respectively.

It is shown in Appendix A that $B_{1}$ is obtained as
\begin{equation}
B_{1}\equiv\frac{24\pi R\,\Phi_{0}}
{7\zeta(3) 
(\chi_{xx}\chi_{yy}\!-\!\chi_{xy}^{2})^{1/2}\,
(\hbar\langle v_{\rm F}^{2}\rangle^{1/2}/T_{c})^{2}}
\, ,
\label{B_1}
\end{equation}
where $\zeta$ is the Riemann zeta function, $\chi_{ij}$ is given by Eq.\ (\ref{chi}),
and $R$ is defined by
\begin{equation}
R = 1-\hbar\frac{1\!-\!\langle\phi\rangle^{2}}{2\tau}
2\pi T_{c}\sum_{n=0}^{\infty} \frac{1}{\tilde{\varepsilon}_{n}^{2}} \, .
\label{R}
\end{equation}
The factor $\hbar\langle v_{\rm F}^{2}\rangle^{1/2}/T_{c}$ in the denominator of
Eq.\ (\ref{B_1}) is essentially the BCS coherence length.\cite{BCS}
Also, $R$ is dimensionless and approaches
unity for $\tau\!\rightarrow\!\infty$.
Equation (\ref{B_1}) is a direct generalization of the result
by Rieck and Scharnberg\cite{RS90} for $\phi({\bf k}_{\rm F})\!=\! 1$ 
to the cases with gap anisotropy and for arbitrary strength of the impurity scattering.

It is convenient to
express $\langle v_{{\rm F}i}v_{{\rm F}j} \rangle$ in Eq.\ (\ref{chi}) 
with respect to the crystallographic coordinates $(X,Y,Z)$
to see the anisotropy in $B_{1}$ manifestly.
Using Eq.\ (\ref{calR}), $v_{{\rm F}x}$ and $v_{{\rm F}y}$ are rewritten as
\begin{equation}
\left\{
	\begin{array}{l}
	\vspace{2mm}
    v_{{\rm F}x}=v_{{\rm F}X}\cos\theta\cos\varphi +v_{{\rm F}Y} \cos\theta\sin\varphi-
    v_{{\rm F}Z}\sin\theta \\
    v_{{\rm F}y}=-v_{{\rm F}X}\sin\varphi +v_{{\rm F}Y}\cos\varphi
	\end{array}
\right. ,
\end{equation}
so that 
\begin{equation}
\left\{
	\begin{array}{l}
    \langle v_{{\rm F}x}^{2}\rangle=(\langle v_{{\rm F}X}^{2}\rangle\cos^{2}\varphi 
    +\langle v_{{\rm F}Y}^{2}\rangle\sin^{2}\varphi)\cos^{2}\theta\\
	\vspace{2mm}
     \hspace{12mm} +
    \langle v_{{\rm F}Z}^{2}\rangle\sin^{2}\theta \\
    \vspace{2mm}
   \langle v_{{\rm F}y}^{2}\rangle =\langle v_{{\rm F}X}^{2}\rangle
    \sin^{2}\varphi + \langle v_{{\rm F}Y}^{2}\rangle\cos^{2}\varphi \\
    \langle v_{{\rm F}x}v_{{\rm F}y}\rangle =(\langle v_{{\rm F}Y}^{2}\rangle
    -\langle v_{{\rm F}X}^{2}\rangle)\cos\theta\cos\varphi\sin\varphi
	\end{array}
\right. .
\label{<vFvF>}
\end{equation}
The quantities $\langle \phi\, v_{{\rm F}x}v_{{\rm F}y}\rangle$ and 
$\langle \phi^{2} v_{{\rm F}x}v_{{\rm F}y}\rangle$ can be expressed similarly
in the crystallographic coordinates
once $\phi({\bf k}_{{\rm F}})$ is given explicitly.
In particular, when $\phi({\bf k}_{{\rm F}})$ belongs to the $A_{1g}$ representation,
the expressions for the two averages are essentially the same as Eq.\ (\ref{<vFvF>}).
From Eqs.\ (\ref{B_1}),  (\ref{chi}), and (\ref{<vFvF>}),
we realize immediately that the initial slope is isotropic when (i)
$\phi({\bf k}_{{\rm F}})$ belongs to $A_{1g}$ and (ii) the crystal has 
cubic symmetry.

The expression for $B_{2}$ is more complicated as given explicitly by Eq.\ (\ref{B_2}).
It includes Fermi-surface averages of $v_{Fx}^{4}$, $v_{Fx}^{2}v_{Fy}^{2}$, etc., 
and enables us to estimate the initial curvature of $H_{c2}$ given the Fermi-surface
structure.

\subsection{$\bm H_{c2}$ equation}

We now derive an $H_{c2}$ equation which can be solved efficiently at all temperatures.
To this end,
we transform Eqs.\ (\ref{Eilen(1)}) and (\ref{pair(1)})
into algebraic equations by expanding
$\Delta$ and $f^{(1)}$ in the basis functions of
of Eq.\ (\ref{basis}) as\cite{Kita98,Kita03}
\begin{subequations}
\label{DfExpand}
\begin{equation}
\Delta({\bf r})= \sqrt{V}\sum_{N=0}^{\infty}\Delta_{N}\,\psi_{N{\bf q}}({\bf r}) \, ,
\label{DExpand}
\end{equation}
\begin{equation}
f^{(1)}(\varepsilon_{n},{\bf k}_{{\rm F}},{\bf r})
= \sqrt{V}\sum_{N=0}^{\infty}f^{(1)}_{N}(\varepsilon_{n},{\bf k}_{{\rm F}})\,
\psi_{N{\bf q}}({\bf r}) \, .
\label{fExpand}
\end{equation}
\end{subequations}
Let us substitute Eqs.\ (\ref{gradient}) and (\ref{DfExpand}) into Eqs.\ (\ref{Eilen(1)}) and 
(\ref{pair(1)}), multiply them by $\psi_{N{\bf q}}^{*}({\bf r})$, and perform
integrations over ${\bf r}$. Equations (\ref{Eilen(1)}) and 
(\ref{pair(1)}) are thereby transformed into
\begin{subequations}
\label{Eilens(1)-2}
\begin{equation}
\sum_{N'} {\cal M}_{NN'}f_{N'}^{(1)} = \phi\Delta_{N}
+\frac{\hbar}{2\tau}\langle {f}_{N}^{(1)}\rangle \, ,
\label{Eilen(1)-2}
\end{equation}
\begin{equation}
\Delta_{N} \ln \frac{T_{c0}}{T}=-\pi T \sum_{n=-\infty}^{\infty}
\left(\langle \phi{f}_{N}^{(1)} \rangle -\frac{\Delta_{N}}{|\varepsilon_{n}|}\right) \, ,
\label{pair(1)-2}
\end{equation}
\end{subequations}
where the matrix ${\cal M}$ is tridiagonal as
\begin{equation}
{\cal M}_{NN'}\equiv\tilde{\varepsilon}_{n}'\delta_{NN'}
+\sqrt{N\!+\!1}\,\bar{\beta}^{*} \delta_{N,N'-1}
- \sqrt{N} \bar{\beta}\,\delta_{N,N'+1}\, ,
\label{Matrix}
\end{equation}
with 
\begin{equation}
\bar{\beta}\equiv\frac{\hbar \bar{v}_{{\rm F}+}{\rm sgn}(\varepsilon_{n})}
{2\sqrt{2}\,l_{{\rm c}}}\, .
\label{beta}
\end{equation}

We first focus on Eq.\ (\ref{Eilen(1)-2}) and introduce the matrix $\cal K$ by
\begin{eqnarray}
{\cal K}_{NN'}\equiv ({\cal M}^{-1})_{NN'} \, ,
\label{calK}
\end{eqnarray}
which necessarily has the same symmetry as ${\cal M}$:\cite{comment3}
\begin{eqnarray}
&&\hspace{-5mm}{\cal K}_{NN'}(\varepsilon_{n},\bar{\beta})
={\cal K}_{N'N}(\varepsilon_{n},-\bar{\beta}^{*})
={\cal K}_{NN'}^{*}(-\varepsilon_{n},-\bar{\beta}^{*})
\nonumber \\
&&\hspace{15mm}
={\cal K}_{N'N}^{*}(-\varepsilon_{n},\bar{\beta}) \, .
\label{calK-sym}
\end{eqnarray}
Using ${\cal K}$, Eq.\ (\ref{Eilen(1)-2}) is solved formally as
\begin{equation}
f^{(1)}_{N}=\sum_{N'}{\cal K}_{NN'}\left(
\phi\Delta_{N'}+\frac{\hbar}{2\tau}\langle {f}_{N'}^{(1)}\rangle\right) \, .
\label{f^(1)0}
\end{equation}
Taking the Fermi-surface average to obtain $\langle {f}_{N}^{(1)}\rangle$
and substituting it back into Eq.\ (\ref{f^(1)0}), we arrive at
an expression for the vector
${\bf f}^{(1)}\!\equiv\!(f^{(1)}_{0},f^{(1)}_{1},f^{(1)}_{2},\cdots)^{\rm T}$ as
\begin{equation}
{\bf f}^{(1)}=\biggl[{\cal K}\phi\!+\!\frac{\hbar}{2\tau}{\cal K}
\left(\!{\cal I}-\frac{\hbar\langle {\cal K}\rangle}{2\tau}\! \right)^{\!\! -1}\!
\langle{\cal K}\phi\rangle\biggr]{\bm \Delta} \, ,
\label{f^(1)1}
\end{equation}
with ${\cal I}$ the unit matrix in the Landau-level indices
and ${\bm \Delta}\!\equiv\!
(\Delta_{0},\Delta_{1},\Delta_{2},\cdots)^{\rm T}$.

We next substitute Eq.\ (\ref{f^(1)1}) into Eq.\ (\ref{pair(1)-2}).
We thereby obtain the condition that Eq.\ (\ref{pair(1)-2}) has
a nontrivial solution for ${\bm \Delta}$ as
\begin{equation}
\det {\cal A} = 0 \, ,
\label{detA}
\end{equation}
where the matrix ${\cal A}$ is defined by
\begin{eqnarray}
&&\hspace{-7mm}{\cal A}={\cal I}\ln \frac{T}{T_{c0}}
+\pi T \sum_{n=-\infty}^{\infty}\biggl[ \frac{{\cal I}}{|\varepsilon_{n}|} 
-\langle {\cal K}\phi^{2}\rangle
\nonumber \\
&&\hspace{12mm}
-\frac{\hbar}{2\tau}\langle {\cal K}\phi\rangle
\left(\!{\cal I}-\frac{\hbar\langle {\cal K}\rangle}{2\tau}\! \right)^{\!\! -1}\!
\langle{\cal K}\phi\rangle \biggr] \, ,
\label{calA}
\end{eqnarray}
with ${\cal I}$ the unit matrix in the Landau-level indices.
The upper critical field $B_{c2}$
corresponds to the highest field where Eq.\ (\ref{detA})
is satisfied, with $B$ and $H$ connected by Eq.\ (\ref{H-B}).
Put it another way, $B_{c2}$ is determined by requiring that
the smallest eigenvalue of ${\cal A}$ be zero.
Notice that ${\cal A}$ is Hermitian, as can be shown by using
Eq.\ (\ref{calK-sym}),
so that it can be diagonalized easily.

Equation (\ref{calA}) tells us that central to determining $B_{c2}$
lies the calculation of ${\cal K}_{NN'}$ defined by Eqs.\ (\ref{Matrix}) and (\ref{calK}).
An efficient algorithm for it was already developed in Sec.\ IIF
of Ref.\ \onlinecite{Kita03}, which is summarized as follows.
Let us define $R_{N}$ $(N\!=\!0,1,2,\cdots)$ 
and $\bar{R}_{N}$ $(N\!=\!1,2,\cdots)$ by 
\begin{subequations}
\label{R_N}
\begin{equation}
R_{N-1}=(1+Nx^{2}R_{N})^{-1} \, ,
\label{R_Na}
\end{equation}
\begin{equation}
\bar{R}_{N+1}=(1+Nx^{2}\bar{R}_{N})^{-1} \, ,
\hspace{5mm} \bar{R}_{1}= 1 \, ,
\label{R_Nb}
\end{equation}
\end{subequations}
respectively, with 
\begin{equation}
x\!\equiv\!|\bar{\beta}|/\tilde{\varepsilon}_{n}' \, .
\label{x-def}
\end{equation}
Then ${\cal K}_{NN'}$ for $N\!\geq \! N'$ can be obtained by
\begin{equation}
{\cal K}_{NN'}= \frac{1 }{\tilde{\varepsilon}_{n}'}\displaystyle\eta_{N}(x) \bar{\eta}_{N'}(x)
\!\left(\frac{\bar{\beta}}{\tilde{\varepsilon}_{n}'}\right)^{N-N'}
\, , 
\label{K_NN'-2}
\end{equation}
with
\begin{subequations}
\begin{equation}
\eta_{N}\equiv \sqrt{N!} \prod_{k=0}^{N} R_{k}\, ,
\label{PN}
\end{equation}
\begin{equation}
\bar{\eta}_{N}\equiv  
\left\{
	\begin{array}{ll}
	\hspace{3mm} 1 & \hspace{2mm}(N=0) \\
	\displaystyle \frac{1}{\sqrt{N!}} \prod_{k=1}^{N} \frac{1}{\bar{R}_{k}} & 
	\hspace{2mm}(N\geq 1) 
	\end{array}
\right. .
\label{BPN}
\end{equation}
\end{subequations}
The expression of ${\cal K}_{NN'}$ for $N\!< \! N'$ follows
immediately by Eq.\ (\ref{calK-sym}).

As shown in Appendix B, Eqs.\ (\ref{PN}) and (\ref{BPN}) can be written alternatively as
\begin{subequations}
\label{P_N-BP_N}
\begin{eqnarray}
&&\hspace{-5mm}\eta_{N}(x) \equiv \frac{2}{\sqrt{\pi N!}}\int_{0}^{\infty}
\frac{s^{N}H_{N}(s)}{1+2x^{2}s^{2}}\,{\rm e}^{-s^{2}}\,ds  
\nonumber \\
&&\hspace{5mm}
= \frac{1}{\sqrt{N!}}\int_{0}^{\infty}\! s^{N}\exp\!\left(\!-s-\frac{x^{2}}{2}s^{2}\right) ds
\nonumber \\
&&\hspace{5mm}
=2^{N}\! \sqrt{\pi N!}\, z^{N+1} {\rm e}^{z^{2}}{\rm i}^{N}{\rm erfc}(z)
\, ,
\label{P_N}
\end{eqnarray}
\begin{eqnarray}
&&\hspace{-10.5mm}
\bar{\eta}_{N}(x) \equiv \frac{1}{\sqrt{N!}}
\left(\frac{x}{\sqrt{2}i}\right)^{\!\! N}\!\! H_{ N}\!
\left(\frac{i}{\sqrt{2}x}\right) 
\nonumber \\
&&\hspace{2mm}
= \left.\frac{1}{y^{N}\!\sqrt{N!}}\,{\rm e}^{-y^{2}/2}
\left(\!\frac{d}{dy}\!\right)^{\!\! N}\!{\rm e}^{\, y^{2}/2}\,\right|_{y=1/x}.
\label{BP_N}
\end{eqnarray}
\end{subequations}
respectively, where $z\!\equiv\! 1/\sqrt{2}x$
and ${\rm i}^{N}{\rm erfc}(z)$ denotes the repeated integral of the error function.\cite{AS72}
The latter function $\bar{\eta}_{N}(x)$ is an $\frac{N}{2}$th-order ($\frac{N\!-\!1}{2}$th-order)
polynomial of $x^{2}$ for $N\!=\!{\rm even}$ (odd).

Thus, the key quantity ${\cal K}_{NN'}$ is given here in a compact separable form 
with respect to $N$ and $N'$.
This is a plausible feature for performing numerical calculations,
which may be considered as one of the main
advantages of the present formalism
over that of Langmann.\cite{Langmann92}
Our ${\cal K}_{00}$ in Eq.\ (\ref{K_NN'-2}) 
is more convenient than Eq.\ (26) of Hohenberg and Werthamer\cite{HW67}
in that $H_{c2}$ near $T_{c}$ is described
in terms of the lowest Landau level for arbitrary crystal structures.

Equations (\ref{detA}) and (\ref{calA}) with Eqs.\ (\ref{K_NN'-2}),
(\ref{varEpsilon}), (\ref{beta}), (\ref{l_c}),
(\ref{v_+}), (\ref{chi}), and (\ref{c_12}) are one of the
main results of the paper (see also Table I).
They enable us efficient calculations of $H_{c2}$ at all temperatures
based on the Fermi surfaces from {\em ab initio} electronic-structure calculations.
They form a direct extension of the Rieck-Scharnberg equation\cite{RS90}
to the cases with gap anisotropy and arbitrary strength of the impurity scattering.
Indeed, Eq.\ (\ref{BP_N}) is written alternatively as
\begin{equation}
\bar{\eta}_{2N}(x) = \frac{1}{\sqrt{(2N)!}\, 2^{N}z^{2N}}P_{N}(2z^{2}) \, ,
\label{eta_RS90}
\end{equation}
with $z\!\equiv\! 1/\sqrt{2}x$,
where $P_{N}$ is the polynomial defined below Eq.\ (6) of Rieck and Scharnberg.\cite{RS90}
Substituting this result and the last expression of Eq.\ (\ref{P_N}) into Eq.\ (\ref{K_NN'-2}),
it can be checked directly that $\tilde{\varepsilon}_{n}{\cal K}_{2N'2N}$ 
for $N'\!\leq\! N$ is equal to $M_{2N'2N}$ in Eq.\ (6)
of Rieck and Scharnberg.\cite{RS90}
Using this fact, one can show that 
the matrix ${\cal A}$ in Eq.\ (\ref{calA})
reduces to the corresponding matrix in Eq.\ (5) of Rieck and Scharnberg
either
(i) for the isotropic gap with arbitrary impurity scattering
or (ii) in the clean limit with an arbitrary gap structure.
Here we have adopted $x$ in Eq.\ (\ref{x-def}) as a variable 
instead of $z$, because $x$ remains finite at finite temperatures.

From Eq.\ (\ref{K_NN'-2}) and the symmetry
$\bar{\beta}\!\rightarrow\!-\bar{\beta}$ for ${\bf v}_{\rm F}\!
\rightarrow\!-{\bf v}_{\rm F}$, we realize that
$\langle {\cal K}_{2N,2N'+1}\rangle$, $\langle {\cal K}_{2N,2N'+1}\phi\rangle$,
and $\langle {\cal K}_{2N,2N'+1}\phi^{2} \rangle$
all vanish in the present case where (i) the system has inversion symmetry and (ii)
$\phi({\bf k}_{\rm F})$ belongs to an even-parity representation.
It hence follows that we only have to consider $N\!=\!{\rm even}$ Landau levels
in the calculation of Eq.\ (\ref{calA}).
To obtain a matrix element of Eq.\ (\ref{calA}), 
we have to perform a Fermi surface integral for each $n$
and perform the summation over $n$, 
which is well within the capacity of modern computers, however.
Actual calculations of the smallest eigenvalue may be performed by taking 
only $N\!\leq\! N_{\rm cut}$ Landau levels into account, 
and the convergence can be checked by increasing $N_{\rm cut}$.
We can put $N_{\rm cut}\!=\! 0$ near $T_{c}$ due to Eq.\ (\ref{c_12}),
and have to increase $N_{\rm cut}$ as the temperature is lowered.
However, excellent convergence is expected 
at all temperatures by choosing $N_{\rm cut}\!\alt\! 20$.

\section{Extensions of the ${\bm H}_{\bm c2}$ Equation}

We extend the $H_{c2}$ equation of Sec.\ II in several directions.

\subsection{${\bm p}$-wave impurity scattering}

We first take $p$-wave impurity scattering into account.
In this case, Eq.\ (\ref{Eilen}) is replaced by
\begin{eqnarray}
&&\hspace{-2mm}\left(\!\varepsilon_{n}\!-\! i\mu_{\rm B}B\!+\!\frac{\hbar}{2\tau}\langle g\rangle
\!+\! \frac{3\hbar }{2\tau_{1}}
\hat{\bf k}_{\rm F}\!\cdot\!\langle\hat{\bf k}_{\rm F}' g \rangle'
\!+\!\frac{1}{2}\hbar{\bf v}_{{\rm F}}\!\cdot\!{\bm \partial}\!\right)\! f
\nonumber \\ 
&& \hspace{-3mm}=
\left(\!\phi \Delta \!+\!\frac{\hbar}{2\tau}\langle f\rangle\!+\!
\frac{3\hbar }{2\tau_{1}}
\hat{\bf k}_{\rm F}\!\cdot\!\langle\hat{\bf k}_{\rm F}' f \rangle'\!\right)\! g \, ,
\label{Eilen-p}
\end{eqnarray}
where $\langle\hat{\bf k}_{\rm F}'g\rangle'\!\equiv\!\langle\hat{\bf k}_{\rm F}'
g(\varepsilon_{n},{\bf k}_{\rm F}',{\bf r})\rangle'$, for example, and
$\hat{\bf k}_{\rm F}\!\equiv\!{\bf k}_{\rm F}/\langle k_{\rm F}^{2}\rangle^{1/2}$.
Notice that $\hat{\bf k}_{\rm F}$ is not a unit vector in general.
Linearizing Eq.\ (\ref{Eilen-p}) with respect to $\Delta$, we obtain
\begin{eqnarray}
&&\hspace{-10mm}\left(\tilde{\varepsilon}_{n}'
+\frac{{\rm sgn}(\varepsilon_{n})}{2}\hbar{\bf v}_{{\rm F}}\!\cdot\!{\bm \partial}\right) f^{(1)}
=
\phi \Delta +\frac{\hbar}{2\tau}\langle f^{(1)}\rangle
\nonumber \\ 
&& \hspace{35mm}+
\frac{3\hbar }{2\tau_{1}}
\hat{\bf k}_{\rm F}\!\cdot\!\langle \hat{\bf k}_{\rm F}'f^{(1)}\rangle' \, ,
\label{Eilen(1)-p}
\end{eqnarray}
with $\tilde{\varepsilon}_{n}'$ defined by Eq.\ (\ref{varEpsilon}).

First of all, we derive expressions for $T_{c}$ at $H\!=\! 0$,
the coefficients $(c_{1},c_{2})$ in Eq.\ (\ref{aa*}), 
and $B_{c2}$ near $T_{c}$ 
up to the first order in $1\!-\!t$,
based on Eq.\ (\ref{Eilen(1)-p}) and following the procedure in Sec.\ \ref{subsec:Hc2-Tc}.
It turns out that we only need a change of the definition of 
$\chi_{ij}$ from Eq.\ (\ref{chi}) into
\begin{eqnarray}
&& \hspace{-3mm}
\chi_{ij}\equiv  
\frac{24(\pi T_{c})^{3}}{7\zeta(3) \langle v_{{\rm F}}^{2}\rangle}
\sum_{n=0}^{\infty} \frac{1}{\tilde{\varepsilon}_{n}^{3}}\biggl[ \biggl<
{v}_{{\rm F}i} \hspace{0.3mm} {v}_{{\rm F}j}\biggl|\phi+\frac{\langle \phi\rangle}
{2\tau\varepsilon_{n}/\hbar}
\biggr|^{2} \,\biggr> 
\nonumber \\
&& \hspace{30mm}
+ \frac{3}{2\tau_{1}\tilde{\varepsilon}_{n}/\hbar}({\cal P}^{\dagger}{\cal Q}^{-1}{\cal P})_{ij}
\biggr] \, ,
\label{chi-p}
\end{eqnarray}
where the matrices ${\cal P}$ and ${\cal Q}$ are defined by
\begin{subequations}
\begin{equation}
{\cal P}_{ij}\equiv \left<\!\left(\phi+\frac{\langle\phi\rangle}{2\tau|\varepsilon_{n}|/\hbar}
\right)\! \hat{k}_{{\rm F}i}v_{{\rm F}j}\right> \, ,
\end{equation}
\begin{equation}
{\cal Q}_{ij}\equiv \delta_{ij}-
\frac{3\hbar }{2\tau_{1}\tilde{\varepsilon}_{n}}
\langle \hat{k}_{{\rm F}i}\hat{k}_{{\rm F}j}\rangle \, .
\end{equation}
\end{subequations}
Then $T_{c}$, $(c_{1},c_{2})$, and $B_{1}$ in Eq.\ (\ref{Hc2}) 
are given by the same equations,
i.e., Eqs.\ (\ref{Tc}), (\ref{c_12}), and (\ref{B_1}), respectively.

Using
Eqs.\ (\ref{gradient}) and (\ref{DfExpand}), 
we next transform Eq.\ (\ref{Eilen(1)-p}) 
into an algebraic equation.
The resulting equation can solved in the same way as Eq.\ (\ref{f^(1)0})
to yield
\begin{equation}
{\bf f}^{(1)}={\cal K}\biggl(\!
\phi{\bm \Delta}+\frac{\hbar}{2\tau}\langle {\bf f}^{(1)}\rangle+
\frac{3\hbar }{2\tau_{1}}
\hat{\bf k}_{\rm F}\!\cdot\!\langle \hat{\bf k}_{\rm F}'{\bf f}^{(1)}\rangle'
\! \biggr) \, ,
\label{f^(1)-p}
\end{equation}
where ${\cal K}$ is given by Eq.\ (\ref{K_NN'-2}).
It is convenient to introduce the quantities:
\begin{equation}
p_{0}\equiv \sqrt{\frac{\hbar}{2\tau}} \, , \hspace{5mm}
p_{j}\equiv \sqrt{\frac{3\hbar}{2\tau_{1}}}\hat{k}_{{\rm F}j} 
\hspace{5mm} (j\!=\! x,y,z)\, .
\end{equation}
Then from Eq.\ (\ref{f^(1)-p}), we obtain self-consistent equations
for $\langle p_{0}^{*} {\bf f}^{(1)}\rangle$ and 
$\langle p_{j}^{*}{\bf f}^{(1)}\rangle$ as
\begin{eqnarray}
\left[\begin{array}{c}
\vspace{1mm}
\langle p_{0}^{*} {\bf f}^{(1)}\rangle\\
\vspace{1mm}
\langle p_{x}^{*} {\bf f}^{(1)}\rangle\\
\vspace{1mm}
\langle p_{y}^{*} {\bf f}^{(1)}\rangle\\
\langle p_{z}^{*} {\bf f}^{(1)}\rangle
\end{array}
\right]
= {\cal W}
\left[\begin{array}{c}
\vspace{1mm}
\langle p_{0}^{*} {\cal K}\phi\rangle{\bm \Delta}\\
\vspace{1mm}
\langle p_{x}^{*} {\cal K}\phi \rangle{\bm \Delta}\\
\vspace{1mm}
\langle p_{y}^{*} {\cal K}\phi \rangle{\bm \Delta}\\
\langle p_{z}^{*} {\cal K}\phi \rangle{\bm \Delta}
\end{array}
\right],
\label{<f(1)-p>}
\end{eqnarray}
where the matrix ${\cal W}$ is defined by
\begin{equation}
{\cal W}\!\equiv\!\!\left[\!
\begin{array}{cccc}
\vspace{2mm}
\displaystyle
 {\cal I}\!\!-\!\langle |p_{0}|^{2}{\cal K}\rangle \!&\!\!
-\langle p_{0}^{*}p_{x} {\cal K}\rangle \!&\!\!
-\langle p_{0}^{*}p_{y} {\cal K}\rangle \!&\!\!
-\langle p_{0}^{*}p_{z} {\cal K}\rangle\\
\vspace{2mm}
-\langle p_{x}^{*}p_{0} {\cal K}\rangle \!&\!\!
{\cal I}\!\!-\!\langle |p_{x}|^{2}{\cal K}\rangle \!&\!\!
-\langle p_{x}^{*}p_{y} {\cal K}\rangle \!&\!\!
-\langle p_{x}^{*}p_{z} {\cal K}\rangle \\
\vspace{2mm}
-\langle p_{y}^{*}p_{0} {\cal K}\rangle \!&\!\!
-\langle p_{y}^{*}p_{x} {\cal K}\rangle \!&\!\!
{\cal I}\!\!-\!\langle |p_{y}|^{2}{\cal K}\rangle \!&\!\!
-\langle p_{y}^{*}p_{z} {\cal K}\rangle \\
\vspace{2mm}
-\langle p_{z}^{*}p_{0} {\cal K}\rangle \!&\!\!
-\langle p_{z}^{*}p_{x} {\cal K}\rangle \!&\!\!
-\langle p_{z}^{*}p_{y} {\cal K}\rangle \!&\!\!
{\cal I}\!\!-\!\langle |p_{z}|^{2}{\cal K}\rangle \\
\end{array}
\!\!\right]^{\!-1} \!\!\! .
\label{calL-p}
\end{equation}
The complex conjugations $^{*}$ in Eqs.\ (\ref{<f(1)-p>}) and (\ref{calL-p})
are not necessary here but for a later convenience.
Notice the symmetry ${\cal W}_{lm}^{*}(\varepsilon_{n},\bar{\beta})\!=\!
{\cal W}_{ml}(-\varepsilon_{n},\bar{\beta})$ in the matrix elements of $\cal W$, 
as seen from Eq.\ (\ref{calK-sym}).
Using Eq.\ (\ref{<f(1)-p>}) in Eq.\ (\ref{f^(1)-p}), we obtain an explicit expression for 
${\bf f}^{(1)}$ as
\begin{equation}
{\bf f}^{(1)}= {\cal K}\phi{\bm \Delta}+
 \bigl[ p_{0}{\cal K} \,\,
 p_{x}{\cal K} \,\, 
 p_{y}{\cal K} \,\, 
 p_{z}{\cal K} \bigr]
{\cal W}\!
\left[\!
\begin{array}{c}
\vspace{1mm}
\langle  p_{0}^{*}{\cal K}\phi\rangle\\
\vspace{1mm} 
\langle  p_{x}^{*}{\cal K}\phi\rangle\\
\vspace{1mm}
\langle  p_{y}^{*}{\cal K}\phi\rangle\\
\langle  p_{z}^{*}{\cal K}\phi\rangle\\
\end{array}\!
\right] {\bm \Delta} \, .
\label{f^(1)-p2}
\end{equation}
Finally, let us substitute Eq.\ (\ref{f^(1)-p2}) into Eq.\ (\ref{pair(1)-2}). 
We thereby find that Eq.\ (\ref{calA}) is replaced by
\begin{eqnarray}
&&\hspace{-4mm}{\cal A}={\cal I}\ln \frac{T}{T_{c0}}
+\pi T \sum_{n=-\infty}^{\infty}\biggl\{\frac{{\cal I}}{|\varepsilon_{n}|}-
\langle {\cal K}\phi^{2}\rangle
\nonumber \\
&&\hspace{0mm}
-\bigl[\langle p_{0} {\cal K}\phi\rangle\,
\langle p_{x} {\cal K}\phi\rangle \, 
\langle p_{y} {\cal K}\phi\rangle \, 
\langle p_{z} {\cal K}\phi\rangle \bigr]
{\cal W}\!
\left[\!
\begin{array}{c}
\vspace{1mm}
\langle  p_{0}^{*}{\cal K}\phi\rangle\\
\vspace{1mm} 
\langle  p_{x}^{*}{\cal K}\phi\rangle\\
\vspace{1mm}
\langle  p_{y}^{*}{\cal K}\phi\rangle\\
\langle  p_{z}^{*}{\cal K}\phi\rangle\\
\end{array}\!
\right]
\biggr\} .
\nonumber \\
\label{calA-p}
\end{eqnarray}
As before, $H_{c2}$ is determined by requiring that
the smallest eigenvalue of Eq.\ (\ref{calA-p}) be zero. 
This ${\cal A}$ is Hermitian, as can be shown by using Eq.\ (\ref{calK-sym}) and
${\cal W}_{lm}^{*}(\varepsilon_{n},\bar{\beta})\!\equiv\!
{\cal W}_{ml}(-\varepsilon_{n},\bar{\beta})$.
Thus, Eq.\ (\ref{calA-p}) can be diagonalized easily.

It is straightforward to extend Eq.\ (\ref{calA-p}) 
to a more general impurity scattering with the ${\bf k}_{\rm F}$-dependent
relaxation time $\tau({\bf k}_{\rm F},{\bf k}_{\rm F}')$. 
To this end, we apply the procedure of Eqs.\ (\ref{EigenFS})-(\ref{V_kk'})
to $1/\tau({\bf k}_{\rm F},{\bf k}_{\rm F}')$ to expand it as
\begin{equation}
\frac{1}{\tau({\bf k}_{\rm F},{\bf k}_{\rm F}')}=\sum_{\Gamma j \gamma} 
\frac{\eta^{(\Gamma j)}_{\gamma}({\bf k}_{\rm F})\,
\eta^{(\Gamma j)*}_{\gamma}({\bf k}_{\rm F})}{\tau^{(\Gamma j)}} 
 \, ,
\end{equation}
where $1/\tau^{(\Gamma j)}$ and $\eta^{(\Gamma j)}_{\gamma}({\bf k}_{\rm F})$
denote an eigenvalue and its eigenfunction, respectively.
We then realize that
\begin{equation}
p^{(\Gamma j)}_{\gamma}\equiv \sqrt{\frac{\hbar}{2\tau^{(\Gamma j)}}}\,
\eta^{(\Gamma j)}_{\gamma}({\bf k}_{\rm F})
\end{equation}
substitutes for $p_0$ and $p_{j}$ in Eq.\ (\ref{calA-p}).

\subsection{Spin-orbit impurity scattering}

It was noticed by Werthamer {\em et al}.\ \cite{WHH66} and Maki\cite{Maki66}
that, for high-field superconducting alloys with short mean free paths, 
Pauli paramagnetism has to be incorporated simultaneously 
with spin-orbit impurity scattering.
They presented a theory valid for $\tau\!\ll\!\tau_{\rm so}$,
where $\tau_{\rm so}$ is spin-orbit scattering time.
It was later generalized by Rieck {\em et al}.\ \cite{RSS91} 
for an arbitrary value of $\tau_{\rm so}$.
This effect can also be taken into account easily in 
the formulation. 

In the presence of spin-orbit impurity scattering, 
Eq.\ (\ref{Eilen}) is replaced by
\begin{eqnarray}
&&\hspace{-2mm}\left(\!\varepsilon_{n}\!-\! i\mu_{\rm B}B\!+\!\frac{\hbar}{2\tau}\langle g\rangle
\!+\! \frac{\hbar c_{\rm so}}{2\tau_{\rm so}}
\langle|\hat{\bf k}_{\rm F}\!\times\!\hat{\bf k}_{\rm F}'|^{2} g \rangle'
\!+\!\frac{1}{2}\hbar{\bf v}_{{\rm F}}\!\cdot\!{\bm \partial}\!\right)\! f
\nonumber \\ 
&& \hspace{-3mm}=
\left(\!\phi \Delta \!+\!\frac{\hbar}{2\tau}\langle f\rangle\!+\!
\frac{\hbar c_{\rm so}}{2\tau_{\rm so}}
\langle|\hat{\bf k}_{\rm F}\!\times\!\hat{\bf k}_{\rm F}'|^{2} f \rangle'\!\right)\! g \, ,
\label{Eilen-SO}
\end{eqnarray}
with $c_{\rm so}\!\equiv\!
1/\langle\langle|\hat{\bf k}_{\rm F}\!\times\!\hat{\bf k}_{\rm F}'|^{2}\rangle'\rangle$.
To simplify the notations and make the argument transparent,
it is useful to introduce the quantities:
\begin{subequations}
\begin{equation}
p_{0}\equiv \sqrt{\frac{\hbar}{2\tau}} \, , \hspace{5mm}
p_{ij}\equiv \sqrt{\frac{\hbar c_{\rm so}}{2\tau_{\rm so}}}\, 
(\hat{k}_{\rm F}^{2}\delta_{ij}\!-\!\hat{k}_{{\rm F}i}\hat{k}_{{\rm F}j}) \, ,
\end{equation}
\begin{equation}
q_{0}\equiv \sqrt{\frac{\hbar}{2\tau}} \, , \hspace{5mm}
q_{ij}\equiv \sqrt{\frac{\hbar c_{\rm so}}{2\tau_{\rm so}}}
\,\hat{k}_{{\rm F}i}\hat{k}_{{\rm F}j}(2-\delta_{ij})\, ,
\end{equation}
\end{subequations}
and the vectors:
\begin{subequations}
\label{pq}
\begin{equation}
{\bf p}\equiv (p_{0},p_{xx},p_{yy},p_{zz},p_{xy},p_{yz},p_{zx})^{\rm T} \, ,
\end{equation}
\begin{equation}
{\bf q}\equiv (q_{0},q_{xx},q_{yy},q_{zz},q_{xy},q_{yz},q_{zx})^{\rm T} \, .
\end{equation}
\end{subequations}
Then Eq.\ (\ref{Eilen-SO}) linearized with respect to $\Delta$
is written in terms of Eq.\ (\ref{pq}) as
\begin{equation}
\left(\tilde{\varepsilon}_{n}'
+\frac{{\rm sgn}(\varepsilon_{n})}{2}\hbar{\bf v}_{{\rm F}}\!\cdot\!{\bm \partial}\right) f^{(1)}
=
\phi \Delta +{\bf p}\cdot \langle {\bf q}\, f^{(1)}\rangle
\, ,
\label{Eilen(1)-SO}
\end{equation}
where $\tilde{\varepsilon}_{n}'$ is defined by
\begin{equation}
\tilde{\varepsilon}_{n}'\equiv 
\tilde{\varepsilon}_{n}-i\mu_{\rm B}B{\rm sgn}(\varepsilon_{n})\, ,
\hspace{5mm}
\tilde{\varepsilon}_{n}\equiv |\varepsilon_{n}|+{\bf p}\cdot \langle {\bf q}\rangle \, .
\label{varEpsilon-SO}
\end{equation}
Notice ${\bf p}\cdot \langle {\bf q}\rangle\!=\! \langle {\bf p}\rangle\cdot{\bf q} $.

It follows from the procedure in Sec.\ \ref{subsec:Hc2-Tc} that
$T_{c}$ at $H\!=\!0$ satisfies
\begin{equation}
\ln\frac{T_{c0}}{T_{c}}= 2\pi T_{c}\sum_{n=0}^{\infty}\left[
\frac{1}{\varepsilon_{n}}-\left<\!\frac{\phi^{2}}
{\tilde{\varepsilon}_{n}}\!\right>-\left<\!\frac{\bf{p}^{\rm T}\phi}
{\tilde{\varepsilon}_{n}}\!\right>\!{\cal Q}^{-1}\!\left<\!\frac{\bf{q}\phi}
{\tilde{\varepsilon}_{n}}\!\right>\right] \, ,
\label{Tc-SO}
\end{equation}
where the matrix ${\cal Q}$ is defined by $(r,s\!=\! 0,xx,\cdots,zx)$
\begin{equation}
{\cal Q}_{rs}=\delta_{rs}-\left<\!\frac{q_{r}p_{s}}
{\tilde{\varepsilon}_{n}}\!\right> \, .
\end{equation}
Also, $\chi_{ij}$ in Eq.\ (\ref{chi}) should be modified into
\begin{eqnarray}
&& \hspace{-6mm}
\chi_{ij}\equiv  
\frac{24(\pi T_{c})^{3}}{7\zeta(3) \langle v_{{\rm F}}^{2}\rangle}
\sum_{n=0}^{\infty} \biggl<\frac{v_{{\rm F}i} \hspace{0.3mm} v_{{\rm F}j}}
{\tilde{\varepsilon}_{n}^{3}}
\biggl[\phi+\biggl<\frac{{\bf p}^{\rm T} \phi}{\tilde{\varepsilon}_{n}}\biggr> 
{\cal Q}^{-1}{\bf q}\biggr]
\nonumber \\
&& \hspace{25mm}
\times\biggl[\phi+{\bf p}^{\rm T}
{\cal Q}^{-1}\biggl<\frac{{\bf q}\, \phi}{\tilde{\varepsilon}_{n}}\biggr> \biggr]
\biggr> \, .
\label{chi-SO}
\end{eqnarray}
Finally, $R$ in Eq.\ (\ref{R}) is replaced by
\begin{eqnarray}
&&\hspace{-4mm}
R=1-2\pi T_{c}\sum_{n=0}^{\infty}\biggl[
\left<\!\frac{\phi^{2}{\bf p}\!\cdot\!\langle{\bf q}\rangle}
{\tilde{\varepsilon}_{n}^{2}}\!\right>
\nonumber \\
&&\hspace{3mm}
+\left<\!\frac{{\bf p}\!\cdot\!\langle{\bf q}\rangle{\bf p}^{\rm T}\phi}
{\tilde{\varepsilon}_{n}^{2}}\!\right>\!{\cal Q}^{-1}\!
\left<\!\frac{{\bf q}\, \phi}
{\tilde{\varepsilon}_{n}}\!\right>
-\varepsilon_{n}\!\left<\!\frac{{\bf p}^{\rm T}\phi}
{\tilde{\varepsilon}_{n}}\!\right>\!{\cal Q}^{-1}\!
\left<\!\frac{{\bf q}\, \phi}
{\tilde{\varepsilon}_{n}^{2}}\!\right>
\nonumber \\
&&\hspace{3mm}
-\varepsilon_{n}\!\left<\!\frac{{\bf p}^{\rm T}\phi}
{\tilde{\varepsilon}_{n}}\!\right>\!{\cal Q}^{-1}\!
\left<\!\frac{{\bf q}^{\rm T}{\bf p}}
{\tilde{\varepsilon}_{n}^{2}}\!\right>\!{\cal Q}^{-1}\!
\left<\!\frac{{\bf q}\, \phi}
{\tilde{\varepsilon}_{n}}\!\right>\biggr] \, .
\end{eqnarray}
With the above modifications,
$T_{c}$, $(c_{1},c_{2})$, and $B_{1}$ in Eq.\ (\ref{Hc2}) are given by 
Eqs.\ (\ref{Tc}), (\ref{c_12}), and (\ref{B_1}), respectively.

We now transform Eq.\ (\ref{Eilen(1)-SO}) into an algebraic equation by using
Eqs.\ (\ref{gradient}) and (\ref{DfExpand}).
The resulting equation can solved in the same way as Eq.\ (\ref{f^(1)0}).
We thereby obtain
\begin{equation}
{\bf f}^{(1)}=
{\cal K}\phi{\bm \Delta}+\sum_{r}p_{r} {\cal K} \langle q_{r} {\bf f}^{(1)}\rangle
\, ,
\label{f^(1)-SO}
\end{equation}
where ${\cal K}$ is given by Eq.\ (\ref{K_NN'-2}) with 
$\tilde{\varepsilon}_{n}'$ replaced by Eq.\ (\ref{varEpsilon-SO}).
From Eq.\ (\ref{f^(1)-SO}), we obtain self-consistent equations
for $\langle q_{0}{\bf f}^{(1)}\rangle$ and 
$\langle q_{ij}{\bf f}^{(1)}\rangle$ as
\begin{eqnarray}
\left[\begin{array}{c}
\vspace{1mm}
\langle q_{0} {\bf f}^{(1)}\rangle\\
\vspace{1mm}
\langle q_{xx}{\bf f}^{(1)}\rangle\\
\vspace{1mm}
\langle q_{yy}{\bf f}^{(1)}\rangle\\
\vspace{1mm}
\langle q_{zz}{\bf f}^{(1)}\rangle\\
\vspace{1mm}
\langle q_{xy}{\bf f}^{(1)}\rangle\\
\vspace{1mm}
\langle q_{yz}{\bf f}^{(1)}\rangle\\
\langle q_{zx}{\bf f}^{(1)}\rangle
\end{array}
\right]
= {\cal W}
\left[\begin{array}{c}
\vspace{1mm}
\langle q_{0} {\cal K}\phi\rangle{\bm \Delta}\\
\vspace{1mm}
\langle q_{xx}{\cal K}\phi\rangle{\bm \Delta}\\
\vspace{1mm}
\langle q_{yy}{\cal K}\phi\rangle{\bm \Delta}\\
\vspace{1mm}
\langle q_{zz}{\cal K}\phi\rangle{\bm \Delta}\\
\vspace{1mm}
\langle q_{xy}{\cal K}\phi\rangle{\bm \Delta}\\
\vspace{1mm}
\langle q_{yz}{\cal K}\phi\rangle{\bm \Delta}\\
\langle q_{zx}{\cal K}\phi\rangle{\bm \Delta}
\end{array}
\right] \, ,
\label{<f(1)-SO>}
\end{eqnarray}
where the matrix ${\cal W}$ is defined by
\begin{equation}
{\cal W}\equiv\!\left[\!
\begin{array}{cccc}
\vspace{2mm}
\displaystyle
 {\cal I}\!-\!\langle q_{0}p_{0}{\cal K}\rangle \!&\!
-\langle q_{0}p_{xx} {\cal K}\rangle \!&\! 
-\langle q_{0}p_{yy} {\cal K}\rangle \!&\! 
\cdots \\
\vspace{2mm}
-\langle q_{xx}p_{0} {\cal K}\rangle \!&\!
{\cal I}\!-\!\langle q_{xx}p_{xx} {\cal K}\rangle \!&\! 
-\langle q_{xx}p_{yy} {\cal K}\rangle \!&\! 
\cdots \\
\vspace{2mm}
-\langle q_{yy}p_{0} {\cal K}\rangle \!&\!
-\langle q_{yy}p_{xx} {\cal K}\rangle \!&\! 
{\cal I}\!-\!\langle q_{yy}p_{yy} {\cal K}\rangle \!&\! 
\cdots \\
\vspace{2mm}
-\langle q_{zz}p_{0} {\cal K}\rangle \!&\!
-\langle q_{zz}p_{xx} {\cal K}\rangle \!&\! 
-\langle q_{zz}p_{yy} {\cal K}\rangle \!&\! 
\cdots \\
\vspace{2mm}
-\langle q_{xy}p_{0} {\cal K}\rangle \!&\!
-\langle q_{xy}p_{xx} {\cal K}\rangle \!&\! 
-\langle q_{xy}p_{yy} {\cal K}\rangle \!&\! 
\cdots \\
\vspace{2mm}
-\langle q_{yz}p_{0} {\cal K}\rangle \!&\!
-\langle q_{yz}p_{xx} {\cal K}\rangle \!&\! 
-\langle q_{yz}p_{yy} {\cal K}\rangle \!&\! 
\cdots \\
\vspace{2mm}
-\langle q_{zx}p_{0} {\cal K}\rangle \!&\!
-\langle q_{zx}p_{xx} {\cal K}\rangle \!&\! 
-\langle q_{zx}p_{yy} {\cal K}\rangle \!&\! 
\cdots 
\end{array}
\!\right]^{-1} \! .
\label{calL-so}
\end{equation}
Using Eq.\ (\ref{<f(1)-SO>}) in Eq.\ (\ref{f^(1)-SO}), 
we obtain an explicit expression for ${\bf f}^{(1)}$ as
\begin{eqnarray}
&&\hspace{-4mm}
{\bf f}^{(1)} = {\cal K}\phi{\bm \Delta}\!+\!\bigl[ p_{0}{\cal K} \,\,
 p_{xx}{\cal K} \,\,
 p_{yy}{\cal K} \,\,
\cdots
\bigr]
{\cal W}\!
\left[\!
\begin{array}{c}
\vspace{1mm}
\langle  q_{0}{\cal K}\phi\rangle\\
\vspace{1mm} 
\langle  q_{xx}{\cal K}\phi\rangle\\
\vspace{1mm}
\langle  q_{yy}{\cal K}\phi\rangle\\
\vdots
\end{array}\!
\right] \!\! {\bm \Delta} 
\nonumber \\
&&\hspace{3mm}
= {\cal K}\phi{\bm \Delta}\!+\!\bigl[ q_{0}{\cal K} \,\,
 q_{xx}{\cal K} \,\,
 q_{yy}{\cal K} \,\,
\cdots
\bigr]
{\cal W}^{\dagger}\!
\left[\!
\begin{array}{c}
\vspace{1mm}
\langle  p_{0}{\cal K}\phi\rangle\\
\vspace{1mm} 
\langle  p_{xx}{\cal K}\phi\rangle\\
\vspace{1mm}
\langle  p_{yy}{\cal K}\phi\rangle\\
\vdots
\end{array}\!
\right] \!\! {\bm \Delta} \, ,
\nonumber \\
\label{f^(1)-SO2}
\end{eqnarray}
with ${\cal W}^{\dagger}$ defined by $[{\cal W}^{\dagger}(\varepsilon_{n},\bar{\beta})]_{lm}
\!\equiv\!
{\cal W}^{*}_{ml}(-\varepsilon_{n},\bar{\beta})$.
The latter expression originates from the self-consistency equations
for $\langle p_{0}{\bf f}^{(1)}\rangle$ and $\langle p_{ij}{\bf f}^{(1)}\rangle$ 
similar to Eq.\ (\ref{<f(1)-SO>}).
Finally, let us substitute Eq.\ (\ref{f^(1)-SO2}) into Eq.\ (\ref{pair(1)-2}). 
We thereby find that Eq.\ (\ref{calA}) is replaced by
\begin{eqnarray}
&&\hspace{-4mm}{\cal A}={\cal I}\ln \frac{T}{T_{c0}}
+\pi T \sum_{n=-\infty}^{\infty}\biggl\{\frac{{\cal I}}{|\varepsilon_{n}|}-
\langle {\cal K}\phi^{2}\rangle
\nonumber \\
&&\hspace{1mm}
-\bigl[\langle p_{0}{\cal K}\phi\rangle \,
\langle p_{xx}{\cal K}\phi\rangle \,
\langle p_{yy}{\cal K}\phi\rangle \,\cdots\bigr]
{\cal W}\!
\left[\!
\begin{array}{c}
\vspace{1mm}
\langle q_{0} {\cal K}\phi\rangle\\
\vspace{1mm}
\langle q_{xx}{\cal K}\phi \rangle\\
\vspace{1mm}
\langle q_{yy}{\cal K}\phi \rangle\\
\cdots
\end{array}\!
\right] \biggr\}
\nonumber \\
&&\hspace{-1mm}
={\cal I}\ln \frac{T}{T_{c0}}
+\pi T \sum_{n=-\infty}^{\infty}\biggl\{\frac{{\cal I}}{|\varepsilon_{n}|}-
\langle {\cal K}\phi^{2}\rangle
\nonumber \\
&&\hspace{1mm}
-\bigl[\langle q_{0}{\cal K}\phi\rangle \,
\langle q_{xx}{\cal K}\phi\rangle \,
\langle q_{yy}{\cal K}\phi\rangle \,\cdots\bigr]
{\cal W}^{\dagger}\!
\left[\!
\begin{array}{c}
\vspace{1mm}
\langle p_{0} {\cal K}\phi\rangle\\
\vspace{1mm}
\langle p_{xx}{\cal K}\phi \rangle\\
\vspace{1mm}
\langle p_{yy}{\cal K}\phi \rangle\\
\cdots
\end{array}\!
\right]\biggr\}
 \, .
\nonumber \\
\label{calA-SO}
\end{eqnarray}
As before, $H_{c2}$ is determined by requiring that
the smallest eigenvalue of Eq.\ (\ref{calA-SO}) be zero.
This ${\cal A}$ is Hermitian, as can be shown by using Eq.\ (\ref{calK-sym}) and
$[{\cal W}^{\dagger}(\varepsilon_{n},\bar{\beta})]_{lm}
\!\equiv\!
{\cal W}^{*}_{ml}(-\varepsilon_{n},\bar{\beta})$, which can be diagonalized easily.

\subsection{Strong electron-phonon interactions}

We finally consider the effects of strong electron-phonon interactions
within the framework of the Eliashberg theory.\cite{Eliashberg60,Allen82}
We adopt the notations used by Allen and B. Mitrovi\'c\cite{Allen82}
except the replacement $Z\Delta\!\rightarrow\!\Delta$.

The Eilenberger equations were extended by Teichler\cite{Teichler75-2}
to include the strong-coupling effects.
They can also be derived directly from
the equations given by Allen and B. Mitrovi\'c\cite{Allen82} 
by carrying out the ``$\xi$ integration''\cite{SR83} as
\begin{subequations}
\label{Eilens-S}
\begin{equation}
\left(Z\varepsilon_{n}\!-\! i\mu_{\rm B}B\!+\!\frac{\hbar}{2\tau}\langle g\rangle
\!+\!\frac{1}{2}\hbar{\bf v}_{{\rm F}}\!\cdot\!{\bm \partial}\right)\! f=
\left(\Delta \phi \!+\!\frac{\hbar}{2\tau}\langle f\rangle\right)\! g \, ,
\label{Eilen-S}
\end{equation}
\begin{equation}
\Delta(\varepsilon_{n},{\bf r}) 
= \pi T \!\!\! \sum_{n'=-n_{c0}}^{n_{c0}}\!\!
[\lambda(\varepsilon_{n}\!-\!\varepsilon_{n'})\!-\!\mu^{*}]
\langle 
\phi({\bf k}_{\rm F})f(\varepsilon_{n'},{\bf k}_{\rm F},{\bf r})\rangle \, ,
\label{pair-S}
\end{equation}
\begin{equation}
Z(\varepsilon_{n},{\bf k}_{{\rm F}}) 
\!=\! 1\! + \!
\frac{\pi T}{\varepsilon_{n}} \!\! \sum_{n'=-n_{c0}}^{n_{c0}}\!\!\!\!
\langle \lambda({\bf k}_{{\rm F}},{\bf k}_{{\rm F}}',\varepsilon_{n}\!-\!\varepsilon_{n'})
g(\varepsilon_{n'},{\bf k}_{\rm F}',{\bf r})\rangle'   ,
\label{Z-S}
\end{equation}
\end{subequations}
where $n_{c0}$ corresponds to the Matsubara frequency
about five times as large as the Debye frequency.\cite{Allen82}
We have retained full ${\bf k}_{{\rm F}}$ dependence of $\lambda$ in Eq.\ (\ref{Z-S}), 
because the contribution from other pairing channels, 
which may be negligible for the pair potential,
can be substantial for the renormalization factor $Z$.

We linearize Eqs.\ (\ref{Eilens-S}) with respect to $\Delta$ 
and repeat the procedure in Sec.\ \ref{subsec:Hc2-Tc}
up to the zeroth order in $1\!-\!t$.
It then follows that $T_{c}$ at $H\!=\!0$ is determined by the condition that
the smallest eigenvalue of the following matrix be zero:
\begin{eqnarray}
&&\hspace{-8mm}
{\cal A}_{nn'}^{(0)}\equiv\delta_{nn'}
-\pi T [\lambda(\varepsilon_{n}\!-\!\varepsilon_{n'})\!-\!\mu^{*}]\biggl[
\biggl< \frac{\phi^{2}}{\tilde{\varepsilon}_{n'}}\biggr>
\nonumber \\
&&\hspace{20mm}
+\frac{\hbar}{2\tau}
\biggl< \frac{\phi}{\tilde{\varepsilon}_{n'}}\biggr>^{\!\! 2}
\biggl< \frac{\tilde{\varepsilon}_{n'}}{Z_{n'}^{(0)}|\varepsilon_{n'}|}\biggr>\biggr]\, ,
\label{calA0-S}
\end{eqnarray}
where $Z^{(0)}$ is given by
\begin{equation}
Z^{(0)}(\varepsilon_{n},{\bf k}_{{\rm F}}) = 1+
\frac{\pi T}{\varepsilon_{n}} \!\! \sum_{n'=-n_{c0}}^{n_{c0}}\!\!\!\!
\langle\lambda({\bf k}_{{\rm F}},{\bf k}_{{\rm F}}',\varepsilon_{n}-\varepsilon_{n'}) \rangle'
{\rm sgn}(\varepsilon_{n'}) 
\, ,
\label{Z(0)-S}
\end{equation}
and $\tilde{\varepsilon}_{n}$ is defined together with $\tilde{\varepsilon}_{n}'$
by
\begin{equation}
\tilde{\varepsilon}_{n}\equiv Z^{(0)}|\varepsilon_{n}|+\frac{\hbar}{2\tau}
\,  , \hspace{5mm}
\tilde{\varepsilon}_{n}'\equiv 
\tilde{\varepsilon}_{n}-i\mu_{\rm B}B{\rm sgn}(\varepsilon_{n}) \, .
\label{varEpsilon-S}
\end{equation}
We next fix $(c_{1},c_{2})$ in Eq.\ (\ref{aa*}) conveniently.
For the weak-coupling model, we have fixed it by using Eq.\ (\ref{PP-aa})
near $T_c$ so that the coefficient of $aa$ vanishes, i.e., there is no
mixing of higher Landau levels in the $H_{c2}$ equation near $T_c$.
However,
the coefficient of $aa$ in the corresponding strong-coupling equation
becomes frequency dependent. 
It hence follows that, even near $T_{c}$,
there is no choice for $(c_{1},c_{2})$ 
which prevents mixing of higher Landau levels from the $H_{c2}$ equation.
We here adopt the weak-coupling expression in Eq.\ (\ref{c_12}).

We now consider the $H_{c2}$ equation
and repeat the same calculations as those in Sec.\ IIB.
We thereby find that Eq.\ (\ref{calA}) is replaced by
\begin{eqnarray}
&&\hspace{-5mm}{\cal A}_{nN,n'N'}=\delta_{nn'}\delta_{NN'}
-\pi T[\lambda(\varepsilon_{n}\!-\!\varepsilon_{n'})\!-\!\mu^{*}]
\biggl[\langle {\cal K}'\phi^{2}\rangle
\nonumber \\
&&\hspace{13.5mm}
+\frac{\hbar}{2\tau}\langle {\cal K}'\phi\rangle
\left(\!{\cal I}-\frac{\hbar\langle {\cal K}'\rangle}{2\tau}\! \right)^{\!\! -1}\!
\langle{\cal K}'\phi\rangle\biggr]_{\! NN'},
\label{calA-S}
\end{eqnarray}
where ${\cal K}'\!\equiv\!{\cal K}(\varepsilon_{n'},\bar{\beta})$ which also has
${\bf k}_{\rm F}$ dependence through $Z^{(0)\prime}\!=\!Z^{(0)}(\varepsilon_{n'},{\bf k}_{\rm F})$.
As before, $H_{c2}$ is determined by requiring that
the smallest eigenvalue of Eq.\ (\ref{calA-S}) be zero. 

We may alternatively use, instead of Eq.\ (\ref{calA-S}), the matrix:
\begin{eqnarray}
&&\hspace{-8mm}{\cal A}_{nN,n'N'}'=
(\lambda\!-\!\mu^{*})^{-1}_{nn'}\delta_{NN'}
-\delta_{nn'}\pi T
\biggl[\langle {\cal K}\phi^{2}\rangle
\nonumber \\
&&\hspace{10.5mm}
+\frac{\hbar}{2\tau}\langle {\cal K}\phi\rangle
\left(\!{\cal I}-\frac{\hbar\langle {\cal K}\rangle}{2\tau}\! \right)^{\!\! -1}\!
\langle{\cal K}\phi\rangle\biggr]_{\! NN'},
\label{calA-S'}
\end{eqnarray}
where $(\lambda\!-\!\mu^{*})^{-1}$ denotes inverse matrix of $\lambda\!-\!\mu^{*}$.
It is Hermitian for $\mu_{\rm B}B\!\rightarrow\!0$, and also acquire the property
by combining $n\!>\!0$ and $n\!<\!0$ elements.

\section{Model Calculations}

\begin{figure}[tb]
  \includegraphics[width=7cm]{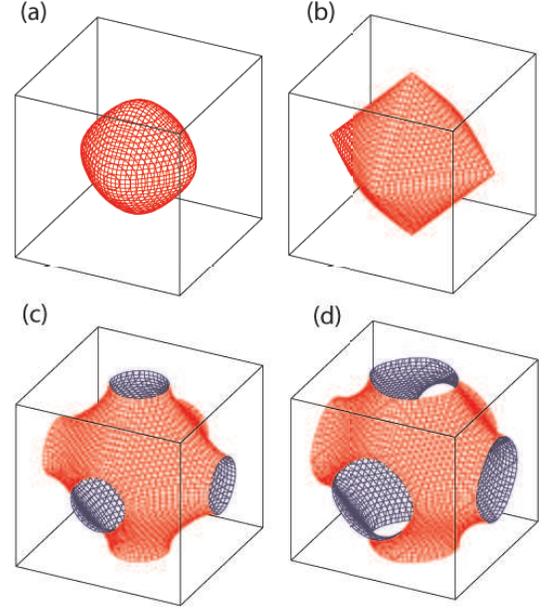}
  \caption{Fermi Surfaces of the tight-binding model in the simple 
  cubic lattice. 
  The Fermi energies are: 
  (a) $\varepsilon_{\rm F} = -3$, (b) $-2$, (c) $-1$, and (d) $0$. }
  \label{fig:tb3dfs}
\end{figure}

We now present results of a model calculation 
based on the formalism developed above.
We restrict ourselves to the weak-coupling model of Sec.\ II with
an isotropic gap, no impurities,
and no Pauli paramagnetism.
As for the energy-band structure, 
we adopt a tight-binding model in the simple cubic lattice
whose dispersion is given by
\begin{equation}
    \varepsilon_{{\bf k}} = -2t \left\{ \cos(k_x a) + \cos(k_y a) + \cos(k_z a)
    \right\}.
    \label{eq:tb3d}
\end{equation}
Here $a$ denotes lattice spacing of the cubic unit cell and 
$t$ is the nearest-neighbor transfer integral.  
We set $t\!=\!a\!=\!1$ in the following.
The corresponding Fermi surfaces are plotted in Fig.~\ref{fig:tb3dfs}
for various values of the Fermi energy $\varepsilon_{\rm F}$. 
For $\varepsilon_{\rm F}\!\approx\!-6$, i.e., near the bottom of the band, 
the Fermi surface is almost spherical with
slight distortion due to the cubic symmetry.
As $\varepsilon_{\rm F}$ increases, the cubic distortion is gradually enhanced. 
Then at $\varepsilon_{\rm F} \!=\! -2$,
 the Fermi surface touches the Brillouin-zone 
boundary at ${\bf k}_{X} \!\equiv\! (0, 0, \pm \pi), (0, \pm, \pi, 0), (\pm \pi, 0, 0)$. 
Above this critical Fermi energy, the topology of the Fermi
surface changes as shown in Fig.~\ref{fig:tb3dfs}(c). 
It is interesting to
see how such a topological change of the Fermi surface affects 
$H_{c2}$.

We computed $H_{c2}$ based on Eq.\ (\ref{detA}) in the clean limit
without Pauli paramagnetism.
The Fermi-surface average in Eq.\ (\ref{calA}) was performed by two different methods.
For general values of $\varepsilon_{\rm F}$, 
we used the linear tetrahedron method which is applicable to any
structure of the Fermi surface. 
In this method, the irreducible Brillouin zone is divided into a collection of small tetrahedra.
From each tetrahedron which intersects the Fermi surface, a segment of the Fermi
surface is obtained as a polygon by a linear interpolation of the energy band.
Numerical integrations over the Fermi surface were then performed as 
a sum over those polygons. 
Another description of the Fermi
surface is possible for $\varepsilon_{\rm F} \!<\! -2$,
where we can adopt the polar
coordinate ${\bf k} \!= \! 
(k \sin\theta\cos\phi, k\sin\theta\sin\phi,k\cos\theta)$
and the Fermi surface $k_{\rm F}\! =\! k_{\rm F}(\theta, \phi)$ 
is obtained by
solving the equation $\varepsilon_{{\bf k}} = \varepsilon_{\rm F}$ numerically for
each $(\theta, \phi)$. An integration over the
Fermi surface is then performed by using the variables
$(\theta,\phi)$.  
We performed both types of calculations to check the
numerical convergence of the tetrahedron method. 
Excellent agreements were achieved generally by using 3375 tetrahedrons.
An exception is the region
$\varepsilon_{\rm F} \approx -2$,
where larger number of tetrahedrons was necessary 
due to the singularity around ${\bf k}_{X}$.

\begin{figure}[tb]
  \includegraphics[width=7.5cm]{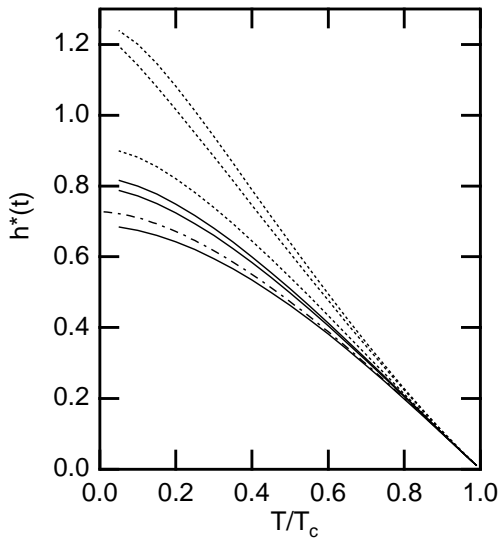}
  \caption{Curves of the reduced critical field $h^*_{d}(t)$ for 
  the cubic tight-binding model with $\varepsilon_{\rm F} \!=\! -2.02$
  (dotted lines), $\varepsilon_{\rm F} \!=\! -3$ (solid lines),
  and $\varepsilon_{\rm F} \!\rightarrow\! -6$ (i.e., the spherical Fermi surface;
  dash-dotted line). The field directions are $d\!=\![111]$, $[110]$, and $[100]$
  from top to bottom in each case.}
  \label{fig:tb3dfig2}
\end{figure}

The infinite matrix ${\cal A}_{NN'}$ in Eq.\ (\ref{calA}) was 
approximated by a finite matrix of $N, N'\!\leq\!N_{\rm cut}$,
and the convergence was checked by increasing $N_{\rm cut}$.
The choice $N_{\rm cut}\!=\!0$ is sufficient
for $T\!\alt T_{c}$, and
it was found numerically that $N_{\rm cut}\!=\!8$ yields
enough convergence for all
field directions at the lowest temperatures.
It was also found that higher Landau levels of $N\!\geq\! 1$
contribute to $H_{c2}$ by only $4$\% even at
$T/T_{c} \!=\! 0.05$.
Thus, the lowest-Landau-level approximation to the pair potential
is excellent for this cubic lattice.
This is not generally the case, however,
and the contribution of higher Landau levels can be considerable
for low-symmetry crystals, as will be reported elsewhere.\cite{Arai04}

Before presenting any detailed results, it is worth noting
that the GL equations,\cite{GL,Abrikosov57}
where the anisotropy enters only through the effective-mass tensor, 
cannot explain possible anisotropy of $H_{c2}$ in cubic
symmetry, as already pointed out by Hohenberg and Werthamer.\cite{HW67}
This GL theory is valid near $T_c$ so that
the upper critical field for $T\!\alt\! T_c$ should be isotropic 
in the present model.
The anisotropy of $H_{c2}$ in cubic symmetry emerges gradually
at lower temperatures, as seen below.

We calculated the reduced critical field 
$h^*(t)$ defined by Eq.\ (\ref{H^*})
for the magnetic field directions $d \! =\! [100]$, $[110]$, and
$[111]$; we denote them as $h^*_{d}(t)$.
Figure~\ref{fig:tb3dfig2} presents
$h^*_{d}(t)$ for $\varepsilon_{\rm F}\! =\! -3$ and $-2.02$ 
as a function of $t \!=\! T/T_c$. 
For $\varepsilon_{\rm F} = -3$, $h^*(t)$ is almost isotropic for
$t \agt 0.8$ and cannot be distinguished from the curve
for the spherical Fermi surface.  
At lower temperatures, the anisotropy appears gradually. 
Whereas $h^*_{[100]}(t)$ 
is reduced from the value for the spherical Fermi surface,
$h^*_{[111]}(t)$ and $h^*_{[110]}(t)$ are enhanced due to the cubic
distortion of the Fermi surface. 
At $t\! =\! 0.05$, $h^*_{[111]}(t)$ and $h^*_{[110]}(t)$ 
are larger than $h^*_{[100]}(t)$ by $19$\% and $15$\%, respectively. 
In another case $\varepsilon_{\rm F}\! =\! -2.02$ 
where the Fermi surface nearly touches the Brillouin zone boundary,
$h^*_{d}(t)$ are remarkably enhanced for all field directions. 
Especially, $h^*_{[111]}(t)$ and
$h^*_{[110]}(t)$ at low temperatures
exhibit values about $60$-$70\%$ 
larger than those for the spherical
Fermi surface.

\begin{table}
  \caption{\label{tab:B2} The ratio $B_2/B_1$ for the field directions
  [100], [110], and [111] in the cases $\varepsilon_{\rm F}\! =\! -3$ and $-2.02$. 
  The quantities $B_1$ and $B_2$ are defined in
  Eq.\ (\ref{Hc2}).
  The values should be compared with 0.13 for the spherical Fermi surface.}
\begin{ruledtabular}
\begin{tabular}{cccc}
  $\varepsilon_F$  & [100] & [110] & [111] \\
  \hline
  $-3$            & 0.08 & 0.27 & 0.33 \\
  $-2.02$         & 0.44 & 0.78 & 0.90 \\
\end{tabular}
\end{ruledtabular}
\end{table}

At $\varepsilon_{\rm F}\! =\! -3$, $h^*_{[111]}(t)$ and $h^*_{[110]}(t)$ near $T_{c}$
show small upward curvature, 
whereas $h^*_{[100]}(t)$ remains almost identical with
the curve for the spherical Fermi surface.  
This difference may be quantified by the ratio
$B_2/B_1$ defined in Eq.\ (\ref{Hc2}). 
It was numerically evaluated by using 
the Fermi velocity on the Fermi surface 
and shown in Table~\ref{tab:B2}. 
The values for the directions
[110] and [111] are larger than $0.13$ for the spherical Fermi surface.
Thus, calculated $B_2/B_1$ values well describe the difference in
$h^*(t)$ for $t\!\alt\! 1$ among field directions.
The upward curvature is more and more pronounced as 
the Fermi surface approaches the Brillouin zone boundary,
as can be seen clearly in Fig.\ 2 for
$\varepsilon_{\rm F}\! =\! -2.02$.
The corresponding ratio $B_2/B_1$ for the [110] and [111] directions are about 
three times larger than those for $\varepsilon_{\rm F}\! = \! -3$.
Thus, the present calculation clearly indicates that 
the Fermi surface anisotropy can be a main source of the upward
curvature in $H_{c2}$ near $T_c$.

\begin{figure}
  \includegraphics[width=6.5cm]{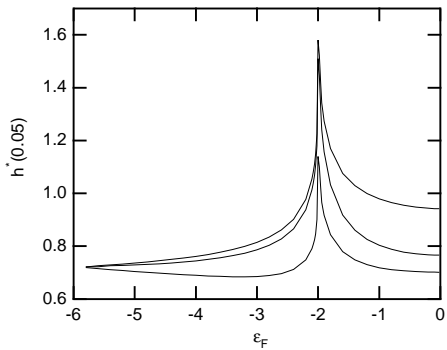}
  \caption{The reduced upper critical field $h^*_d(t)$ at $t = 0.05$ 
  as a function of the Fermi energy $\varepsilon_{\rm F}$. 
  The field directions 
  are $d\!=\![111]$, [110], and [100] from top to bottom, respectively.}
  \label{fig:tb3dfig3}
\end{figure}

In Fig.~\ref{fig:tb3dfig3}, we plot $h^*_{d}(t)$ at $t\!=\!0.05$ as a function of
$\varepsilon_{\rm F}$. As $\varepsilon_{\rm F}\! \rightarrow\! -6$, the
angle dependence of
$h^*_{d}(t)$ vanishes and it converges to the value for the spherical Fermi
surface.  As $\varepsilon_{\rm F}$ is increased from $-6$, cubic distortion is gradually
introduced to the Fermi surface as shown in Fig.\ \ref{fig:tb3dfs},
and $h^*_{d}(t)$ gradually develops anisotropy as a consequence.  For $-6 \!<\!
\varepsilon_{\rm F}\! \lesssim\! -2.5$,  curves of $h^*_{[100]}(t)$ fall below that for
the spherical Fermi surface, whereas $h^*_{[110]}(t)$ and $h^*_{[111]}(t)$ are enhanced
over it.
As $\varepsilon_{\rm F}$ approaches to $-2$, $h^*_d(t)$ is enhanced significantly
irrespective of the field direction. 
Indeed, $h^*_d(t)$ for every field direction shows a singularity 
at $\varepsilon_{\rm F}\!=\!-2$
where the Fermi surface touches the Brillouin zone at ${\bf k}_X$
with vanishing Fermi velocity ${\bf v}_{\rm F}$  at
these points.  As a result, the contribution around these points 
becomes important in the integration $\langle K_{NN'}\rangle$ over the Fermi surface at low
temperatures. This is the origin of the enhancement of $h^*_d(t)$ around
$\varepsilon_{\rm F}\! =\! -2$.  
For $\varepsilon_{\rm F}\! >\! -2$, the difference between $h^*_{[110]}$ 
and $h^*_{[111]}$ 
is larger than that for $\varepsilon_{\rm F}\! \lesssim\! -2.5$. This may be attributed
to the topological difference of the Fermi surface. 
At $\varepsilon_{\rm F}\!=\!0$, the tight-binding band is half-filled and the 
Fermi-surface nesting occurs. However, $h^*_d(t)$ does not show any singularity
around this energy.

\begin{figure}
\includegraphics[width=0.9\linewidth]{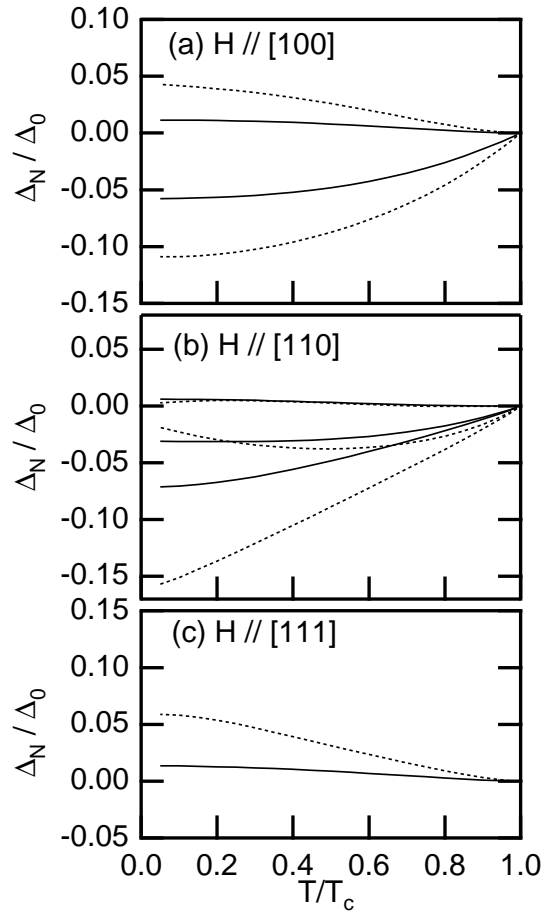}
\caption{The ratio $\Delta_N / 
\Delta_0$ of the expansion coefficients 
in Eq.\ (\ref{DExpand}) as a function of temperature
with (a) ${\bf H}\!\parallel\![100]$, (b) ${\bf H}\!\parallel\![110]$, 
and (c) ${\bf H}\!\parallel\![111]$.
The solid and dotted lines correspond to $\varepsilon_{\rm F} \!=\! -3$ 
and $\varepsilon_{\rm F} \!=\! -2.02$, respectively, with
(a) $N \!=\! 4, 8$ from
bottom to top (b) $N \!=\! 2, 4, 6$ from
bottom to top, and (c) $N \!=\! 6$.
}
\label{fig:tb3dvecfig4}
\end{figure}

Finally, we present results on
the higher Landau-level contributions to the pair potential
$\Delta ({\bf r})$ which is 
expanded as Eq.\ (\ref{DExpand}).
In general, when the system has $n$-fold symmetry around
the field direction, mixing of higher Landau levels 
with multiples of $n$ develops 
as the temperature is lowered.\cite{Kita98}
Figure \ref{fig:tb3dvecfig4} shows the ratio
$\Delta_N / \Delta_0$ as a function of $T/T_{c}$
for $\varepsilon_{\rm F} \!=\! -3$ (solid lines)
and $\varepsilon_{\rm F} \!=\! -2.02$ (dotted lines)
with (a) ${\bf H}\!\parallel\![100]$ ($N \!=\! 4, 8$ from
bottom to top lines), (b) ${\bf H}\!\parallel\![110]$ ($N \!=\! 2, 4, 6$ from
bottom to top lines), and (c) ${\bf H}\!\parallel\![111]$ ($N\!=\! 6$).
One can clearly
observe a general tendency that the mixing is 
more pronounced as
the symmetry around ${\bf H}$ becomes lower as well as
$\varepsilon_{\rm F}$ approaches closer to $-2$.
Especially when ${\bf H}\!\parallel\! [110]$ and
$\varepsilon_{\rm F} \!=\! -2.02$, 
the $N \!=\! 2$ contribution reaches up to nearly $15\%$
of the lowest Landau-level contribution
as $T\!\rightarrow\! 0$.
The results suggest that the lowest-Landau-level approximation for
the pair potential\cite{HW67} is not quantitatively reliable
at low temperatures
for the field along low-symmetry directions,
for complicated Fermi surfaces with divergences in the components
of ${\bf v}_{\rm F}$ perpendicular to ${\bf H}$,
or for low-symmetry crystals.

\section{Summary}

We have derived an efficient $H_{c2}$ equation incorporating
Fermi-surface anisotropy, gap anisotropy, and impurity scattering 
simultaneously.
Basic results of Sec.\ II are summarized in Table I.
This $H_{c2}$ equation is a direct extension of
the Rieck-Scharnberg equation\cite{RS90} and reduces to the latter
either
(i) for the isotropic gap with arbitrary impurity scattering
or (ii) in the clean limit with an arbitrary gap structure,
as shown around Eq.\ (\ref{eta_RS90}).
The operators introduced in Eq.\ (\ref{aa*}) have been helpful to
make the derivation simpler than that by Rieck {\em et al}.\cite{RS90,RSS91}
The present method will be more suitable for extending the consideration
to multi-component-order-parameter systems
or to fields below $H_{c2}$.

We have also obtained a couple of
analytic expressions near $T_{c}$ (i) for $H_{c2}$ 
up to the second order in $1\!-\!T/T_c$ 
and (ii) for the pair potential
up to the first order in $1\!-\!T/T_c$.
The latter result is given by Eq.\ (\ref{DExpandA}) with Eqs.\ (\ref{r_24})
and (\ref{Ws}).
They are useful to estimate the initial curvature of $H_{c2}$
as well as the mixing of higher Landau levels in the pair potential.

The $H_{c2}$ equation of Sec.\ II has also been extended in Sec.\ III to 
include $p$-wave impurity scattering, spin-orbit impurity scattering, 
and strong electron-phonon interactions.

Finally, we have presented numerical examples in Sec.\ IV 
performed for model Fermi surfaces from the 
three-dimensional tight-binding model.
The results clearly demonstrate crucial importance of including 
detailed Fermi-surface structures in the calculation of $H_{c2}$.
It has been found that,
as the Fermi surface approaches the Brillouin zone boundary,
the reduced critical field $h^*(t)$ in Eq.\ (\ref{H^*})
is much enhanced over the value for the isotropic model
with a significant upward curvature near $T_c$.

It is very interesting to see to what degree the upper critical field
of classic type-II superconductors can be described quantitatively
by calculations using realistic Fermi surfaces.
The result by Butler\cite{Butler80-1,Butler80-2} on high-purity Niobium
provides promise to this issue.
We have performed detailed evaluations of $H_{c2}$ for various materials
based on Eq.\ (\ref{detA}) by using Fermi surfaces from
density-functional electronic-structure calculations as an input.
The results are reported elsewhere.\cite{Arai04}

\begin{acknowledgments}
This research is supported by a Grant-in-Aid for Scientific Research 
from the Ministry of Education, Culture, Sports, Science, and Technology
of Japan.
\end{acknowledgments}

\appendix
\section{\label{subsec:Hc2-Tc}Determination of $\bm{(c_{1},c_{2})}$
and analytic Expression of $\bm{H_{c2}}$
near $\bm{T_{c}}$}

We here fix the constants $(c_{1},c_{2})$ in Eqs.\ (\ref{aa*})-(\ref{v_+})
conveniently so that $H_{c2}$ near $T_{c}$ can be described 
in terms of the lowest Landau level only. 
We also derive analytic expressions
for $B_{1}$ and $B_{2}$ in Eq.\ (\ref{Hc2})
so that one can calculate them once the relevant
Fermi-surface structure is given.

In the region $T\!\lesssim\! T_{c}$
where $l_{c}\!\rightarrow\!\infty$ in Eq.\ (\ref{gradient}), 
we can perform a perturbation expansion with respect to 
the gradient operator ${\bf v}_{{\rm F}}\!\cdot\!{\bm \partial}$.
The equation for the $\nu$th-order solution $f^{(1)}_{\nu}$
($\nu\!=\!0,1,\cdots$) is obtained from Eq.\ (\ref{Eilen(1)}) as
\begin{equation}
f^{(1)}_{\nu}=
\delta_{\nu 0}\frac{\phi \Delta}{\tilde{\varepsilon}_{n}'}  
+\frac{\hbar\langle f^{(1)}_{\nu}\rangle}{2\tau\tilde{\varepsilon}_{n}'} 
-\frac{{\rm sgn}(\varepsilon_{n})}{2\tilde{\varepsilon}_{n}'}
\hbar{\bf v}_{{\rm F}}\!\cdot\!{\bm \partial}f^{(1)}_{\nu-1}
\, ,
\label{Eilen(1)-nu}
\end{equation}
with $f^{(1)}_{-1}\!=\!0$.
Noting $\phi(-{\bf k}_{{\rm F}})\!=\!\phi({\bf k}_{{\rm F}})$,
we solve Eq.\ (\ref{Eilen(1)-nu}) self-consistently for $\langle f^{(1)}_{\nu}\rangle$,
put the resulting expression back into Eq.\ (\ref{Eilen(1)-nu}) to 
express $f^{(1)}_{\nu}$ explicitly,
and finally take the Fermi-surface average 
$\langle \phi f^{(1)}_{\nu}\rangle$.
This procedure yields
\begin{subequations}
\label{Eilen(1)_024}
\begin{equation}
\langle \phi f^{(1)}_{0}\rangle =
\frac{1}{\tilde{\varepsilon}_{n}'} \!
\left(\langle \phi^{2}\rangle 
+\frac{\hbar\langle\phi\rangle^{2}}{2\tau|{\varepsilon}_{n}|'}\right)\Delta
\, ,
\label{Eilen(1)_0}
\end{equation}
\begin{equation}
\langle \phi f^{(1)}_{2}\rangle =
\frac{1}{4\tilde{\varepsilon}_{n}^{\prime 3}} \!
\left<\! \left(\phi+\frac{\hbar\langle\phi\rangle}{2\tau|{\varepsilon}_{n}|'}\right)^{\!\! 2}\!
 (\hbar{\bf v}_{{\rm F}}\cdot{\bm \partial})^{2}\right>\Delta  
\, ,
\label{Eilen(1)_2}
\end{equation}
\begin{eqnarray}
&& \hspace{-13mm}
\langle \phi f^{(1)}_{4}\rangle =
\frac{1}{16\tilde{\varepsilon}_{n}^{\prime 5}} \!
\left[\!
\left<\! \left(\phi+\frac{\hbar\langle\phi\rangle}{2\tau|{\varepsilon}_{n}|'}\right)^{\!\! 2}\!
(\hbar{\bf v}_{{\rm F}}\cdot{\bm \partial})^{4}\right> 
\right.
\nonumber \\
&& \hspace{1mm}
\left.  + \frac{\hbar}{2\tau|{\varepsilon}_{n}|'}\!
\left<\!\! \left(\phi+\frac{\hbar\langle\phi\rangle}{2\tau|{\varepsilon}_{n}|'}\right)\!
 (\hbar{\bf v}_{{\rm F}}\cdot{\bm \partial})^{2}\!\right>^{\!\! 2}
 \right]\! \Delta 
\, ,
\label{Eilen(1)_4}
\end{eqnarray}
\end{subequations}
with $|{\varepsilon}_{n}|'\!\equiv\!
|{\varepsilon}_{n}|\!-\!i\mu_{\rm B}B{\rm sgn}(\varepsilon_{n})$, and  $\langle \phi f^{(1)}_{1}\rangle\!=\!\langle \phi f^{(1)}_{3}\rangle\!=\!0$.

Let us substitute Eq.\ (\ref{Eilen(1)_024}) into Eq.\ (\ref{pair(1)}),
replace the gradient operator
by the right-hand side of Eq.\ (\ref{gradient}), put $B\!=\! B_{c2}$ in
$l_{c}$ of Eq.\ (\ref{l_c}), and 
expand $|{\varepsilon}_{n}|'$
with respect to $\mu_{\rm B}B_{c2}/|{\varepsilon}_{n}|$.
We thereby obtain the self-consistency equation near $T_{c}$ as
\begin{eqnarray}
&&\hspace{-3mm}
w_{0,0}\Delta+\frac{B_{c2}}{B_{1}}
\left[w_{2,2}\,a^{\dagger}a^{\dagger}+w_{2,2}^{*}\,aa-w_{2,0}(aa^{\dagger}\!+\! a^{\dagger}a)
\right]\!
\Delta
\nonumber \\
&&\hspace{-3mm}
+\left(\!\frac{B_{c2}}{B_{1}}\!\right)^{\!\!2}
\left[w_{4,4}\,a^{\dagger}a^{\dagger}a^{\dagger}a^{\dagger}+w_{4,4}^{*}\,aaaa
\right.
\nonumber \\
&&\hspace{7mm}
-w_{4,2}\!\left(aa^{\dagger}a^{\dagger}a^{\dagger}+a^{\dagger}aa^{\dagger}a^{\dagger}+
a^{\dagger}a^{\dagger}aa^{\dagger}+a^{\dagger}a^{\dagger}a^{\dagger}a\right)
\nonumber \\
&&\hspace{7mm}
-w_{4,2}^{*}\!\left(a^{\dagger}aaa+aa^{\dagger}aa+
aaa^{\dagger}a+aaaa^{\dagger}\right)
\nonumber \\
&&\hspace{7mm}
+w_{4,0a}\!\left(aa^{\dagger}aa^{\dagger}+a^{\dagger}aaa^{\dagger}+
aa^{\dagger}a^{\dagger}a+a^{\dagger}aa^{\dagger}a\right)
\nonumber \\
&&\hspace{6.6mm}
\left.
+w_{4,0b}\!\left(aaa^{\dagger}a^{\dagger}+a^{\dagger}a^{\dagger}aa\right)
+w_{\rm P}\right]\!\Delta=0 \, .
\label{pairA}
\end{eqnarray}
Here $B_{1}$ is given in Eq.\ (\ref{Hc2}), which is incorporated
into the denominator for convenience.
The functions
$w_{\nu,\mu}\!=\! w_{\nu,\mu}(T)$ and $w_{\rm P}\!=\!w_{\rm P}(T)$ are 
dimensionless  and defined by
\begin{subequations}
\label{Ws}
\begin{equation}
w_{0,0}(T)\equiv \ln\frac{T_{c0}}{T}-\left(1\!-\! \langle\phi\rangle^{2}\right)
2\pi T \sum_{n=0}^{\infty}\!\left(\frac{1}{\varepsilon_{n}}-
\frac{1}{\tilde{\varepsilon}_{n}}\right) \, ,
\label{w00}
\end{equation}
\begin{equation}
w_{2,2}(T)\equiv
\frac{B_{1}\hbar^{2} \pi^{2} T}{2\Phi_{0}} 
\sum_{n=0}^{\infty}\frac{1}{\tilde{\varepsilon}_{n}^{3}}\!
\left<
\!\left(\phi+
\frac{\hbar\langle\phi\rangle}{2\tau\varepsilon_{n}}\right)^{\!\! 2}
\bar{v}_{{\rm F}+}^{2}\right> \, ,
\label{w22}
\end{equation}
\begin{equation}
w_{2,0}(T)\equiv \frac{B_{1}\hbar^{2} \pi^{2} T}{2\Phi_{0}} 
\sum_{n=0}^{\infty}\frac{1}{\tilde{\varepsilon}_{n}^{3}}\!
\left<
\!\left(\phi+
\frac{\hbar\langle\phi\rangle}{2\tau\varepsilon_{n}}\right)^{\!\! 2}
|\bar{v}_{{\rm F}+}|^{2}\right> \, ,
\label{w20}
\end{equation}
\begin{eqnarray}
&&\hspace{-7mm}
w_{4,4}(T)\equiv \frac{B_{1}^{2}\hbar^{4} \pi^{3} T}{8\Phi_{0}^{2}}  
\sum_{n=0}^{\infty}\frac{1}{\tilde{\varepsilon}_{n}^{5}}\!\left[\!
\left<
\!\left(\phi+
\frac{\hbar\langle\phi\rangle}{2\tau\varepsilon_{n}}\right)^{\!\! 2}
\bar{v}_{{\rm F}+}^{4}\right> \right.
\nonumber \\
&&\hspace{15mm}
\left. 
+\frac{\hbar}{2\tau\varepsilon_{n}}\!\left<
\!\left(\phi+
\frac{\hbar\langle\phi\rangle}{2\tau\varepsilon_{n}}\right)
\bar{v}_{{\rm F}+}^{2}\right>^{\!\! 2} \,\right]
\, ,
\label{w44}
\end{eqnarray}
\begin{eqnarray}
&&\hspace{-4mm}
w_{4,2}(T)\equiv \frac{B_{1}^{2}\hbar^{4} \pi^{3} T}{8\Phi_{0}^{2}}  
\sum_{n=0}^{\infty}\frac{1}{\tilde{\varepsilon}_{n}^{5}}\!\left[\!
\left<
\!\left(\phi+
\frac{\hbar\langle\phi\rangle}{2\tau\varepsilon_{n}}\right)^{\!\! 2}
\bar{v}_{{\rm F}+}^{2}|\bar{v}_{{\rm F}+}^{2}|\right> \right.
\nonumber \\
&&\hspace{0mm}
\left. 
+\frac{\hbar}{2\tau\varepsilon_{n}}\!\left<
\!\left(\phi+
\frac{\hbar\langle\phi\rangle}{2\tau\varepsilon_{n}}\right)
\bar{v}_{{\rm F}+}^{2}\right>\!
\left<
\!\left(\phi+
\frac{\hbar\langle\phi\rangle}{2\tau\varepsilon_{n}}\right)
|\bar{v}_{{\rm F}+}|^{2}\right> \right]
\, ,
\nonumber \\
\label{w42}
\end{eqnarray}
\begin{eqnarray}
&&\hspace{-7mm}
w_{4,0a}(T)\equiv \frac{B_{1}^{2}\hbar^{4} \pi^{3} T}{8\Phi_{0}^{2}}  
\sum_{n=0}^{\infty}\frac{1}{\tilde{\varepsilon}_{n}^{5}}\!\left[\!
\left<
\!\left(\phi+
\frac{\hbar\langle\phi\rangle}{2\tau\varepsilon_{n}}\right)^{\!\! 2}
|\bar{v}_{{\rm F}+}|^{4}\right> \right.
\nonumber \\
&&\hspace{15mm}
\left. 
+\frac{\hbar}{2\tau\varepsilon_{n}}\!\left<
\!\left(\phi+
\frac{\hbar\langle\phi\rangle}{2\tau\varepsilon_{n}}\right)
|\bar{v}_{{\rm F}+}|^{2}\right>^{\!\! 2} \,\right]
\, ,
\label{w40a}
\end{eqnarray}
\begin{eqnarray}
&&\hspace{-7mm}
w_{4,0b}(T)\equiv \frac{B_{1}^{2}\hbar^{4} \pi^{3} T}{8\Phi_{0}^{2}}  
\sum_{n=0}^{\infty}\frac{1}{\tilde{\varepsilon}_{n}^{5}}\!\left[\!
\left<
\!\left(\phi+
\frac{\hbar\langle\phi\rangle}{2\tau\varepsilon_{n}}\right)^{\!\! 2}
|\bar{v}_{{\rm F}+}|^{4}\right> \right.
\nonumber \\
&&\hspace{15mm}
\left. 
+\frac{\hbar}{2\tau\varepsilon_{n}}\!\left|\left<
\!\left(\phi+
\frac{\hbar\langle\phi\rangle}{2\tau\varepsilon_{n}}\right)
\bar{v}_{{\rm F}+}^{2}\right>\right|^{2} \,\right]
\, ,
\label{w40b}
\end{eqnarray}
\begin{equation}
w_{\rm P}(T)\equiv - (\mu_{\rm B}B_{1})^{2}\, 
2\pi T \sum_{n=0}^{\infty}
\left[
\frac{\langle\phi\rangle^{2} }{\varepsilon_{n}^{3}}+
\frac{1\!-\! \langle\phi\rangle^{2}}{\tilde{\varepsilon}_{n}^{3}} \right]\, .
\label{w_P}
\end{equation}
\end{subequations}

We next substitute Eq.\ (\ref{Hc2}) into Eq.\ (\ref{pairA}) and
expand $w_{\nu,\mu}$ in Eq.\ (\ref{pairA}) up to the $\frac{4-\nu}{2}$th 
order in $1\!-\!t$. We also put $w_{\rm P}(T)=w_{\rm P}(T_{c})$.
This procedure yields three equations corresponding to 
order $1$, $1\!-\! t$, and $(1\!-\! t)^{2}$.
The equation of order $1$ is given by
$w_{0,0}(T_{c})\Delta \!=\!0$.
It determines $T_{c}$ at $H\!=\! 0$ by
\begin{equation}
\ln\frac{T_{c0}}{T_{c}}= (1\!-\!\langle\phi\rangle^{2})
\left[\psi\!\left(\!\frac{1}{2}\!+\! \frac{\hbar}{4\pi\tau T_{c}}\!\right)
\!-\!\psi\!\left(\frac{1}{2}\right) 
\right] ,
\label{Tc}
\end{equation}
with $\psi(x)$ the digamma function.

The equation of order $1\!-\!t$ in Eq.\ (\ref{pairA}) is obtained as
\begin{eqnarray}
&& \hspace{-8mm}
\bigl[-T_{c}w_{0,0}^{\prime}(T_{c})
-w_{2,0}(T_{c})(2a^{\dagger}a\!+\!1)
\nonumber \\ 
&& \hspace{1mm}
+w_{2,2}(T_{c})a^{\dagger}a^{\dagger}
+w_{2,2}^{*}(T_{c})aa
\bigr]\Delta({\bf r})=0 \, .
\label{PP-aa}
\end{eqnarray}
To solve it, we use the arbitrariness
in $(c_{1},c_{2})$ and impose $w_{2,2}(T_{c})\!=\!0$.
Noting Eqs.\ (\ref{w22}) and (\ref{v_+}), 
this condition is transformed into a dimensionless form as
\begin{equation}
\chi_{xx}c_{2}^{2}+2i\chi_{xy}c_{1}c_{2}-\chi_{yy}c_{1}^{2}=0 \, ,
\label{chi-eq}
\end{equation}
where $\chi_{ij}\!=\!\chi_{ij}(T_{c})$ is defined by Eq.\ (\ref{chi}).
Equation (\ref{chi-eq}) can be solved easily in terms of $c_{2}$.
Substituting the resulting expression into Eq.\ (\ref{c_12-const})
and choosing $c_{1}$ real, we obtain Eq.\ (\ref{c_12}).

Now that $w_{2,2}(T_{c})\!=\!0$ in Eq.\ (\ref{PP-aa}),
the highest field for a nontrivial solution corresponds 
to the lowest Landau level where
$w_{2,0}(T_{c})\!=\!-T_{c}w_{0,0}^{\prime}(T_{c})$.
Introducing 
$R\!\equiv \! -T_{c}w_{0,0}'(T_{c})$ which is given explicitly as Eq.\ (\ref{R}),
and using Eqs.\ (\ref{w00}), (\ref{w20}), (\ref{v_+}), and (\ref{c_12}), 
we obtain the expression for $B_{1}$ as Eq.\ (\ref{B_1}).

We finally consider the equation of order $(1\!-\!t)^{2}$ in Eq.\ (\ref{pairA})
and expand the pair potential as
\begin{equation}
\Delta({\bf r})=\Delta_{0}\bigl\{ \psi_{0{\bf q}}({\bf r})+(1- t)
\bigl[
r_{2}\psi_{2{\bf q}}({\bf r})+r_{4}\psi_{4{\bf q}}({\bf r})\bigr]
\bigr\} \, ,
\label{DExpandA}
\end{equation}
where $\psi_{N{\bf q}}({\bf r})$ is defined by Eq.\ (\ref{basis}),
and $(\Delta_{0},r_{2},r_{4})$ are the expansion coefficients with
$(r_{2},r_{4})$
describing relative mixing 
of higher Landau levels in the pair potential.
Let us substitute Eq.\ (\ref{DExpandA}) into Eq.\ (\ref{pairA}),
multiply the equation of order $(1\!-\!t)^{2}$
by $\psi_{N{\bf q}}^{*}({\bf r})$,
and perform integration over ${\bf r}$.
The resulting equations for $N\!=\! 0,2,4$ yield
\begin{subequations}
\label{B_2-r_24}
\begin{equation}
B_{2}=\frac{\frac{1}{2}T_{c}^{2}w_{0,0}^{(2)}\!+\!T_{c}w_{2,0}'
\!+\!w_{4,0a}\!+\! 2w_{4,0b}\!+\! w_{\rm P}}{R}B_{1} \, ,
\label{B_2}
\end{equation}
\begin{equation}
r_{2}=-\frac{T_{c}w_{2,2}'\!+\! 6w_{4,2}}{2\sqrt{2}\,R} \, ,
\hspace{5mm} r_{4}=\frac{\sqrt{6}\,w_{4,4}}{4R} \, ,
\label{r_24}
\end{equation}
\end{subequations}
respectively.
The functions in Eqs.\ (\ref{B_2}) and (\ref{r_24}) 
are defined by Eqs.\ (\ref{Ws}) and (\ref{R}) and should be evaluated at $T_c$.
In the clean limit $\tau\!\rightarrow\!\infty$, 
these functions acquire simple expressions as
\begin{subequations}
\label{Wsc}
\begin{equation}
R=T_{c}^{2}w_{0,0}^{(2)}=1\, ,\hspace{3mm}  T_{c}w_{2,0}' = -2 \, ,
\hspace{3mm}  T_{c}w_{2,2}' = 0 \, ,
\label{w_00''c}
\end{equation}
\begin{equation}
w_{4,\mu}=\frac{31\zeta(5)}
{[7\zeta(3)]^{2}}
\,\frac{\langle\phi^{2}|\bar{v}_{{\rm F}+}|^{4-\mu}\,\bar{v}_{{\rm F}+}^{\,\mu}
\rangle}
{\langle\phi^{2}|\bar{v}_{{\rm F}+}|^{2}\rangle^{2}} \, ,
\label{w_40c}
\end{equation}
\begin{equation}
w_{\rm P}=-\frac{7\zeta({3})(\mu_{\rm B}B_{1})^{2}}{
4(\pi T_{c})^{2}} \, ,
\label{w_Pc}
\end{equation}
\end{subequations}
with $\mu\!=\!0,2,4$ and $w_{4,0}\!\equiv\!w_{4,0a}\!=\!w_{4,0b}$.
Equation (\ref{B_2-r_24}) with Eq.\ (\ref{Wsc}) includes
the result by Hohenberg and Werthamer\cite{HW67} for cubic materials,
and also the one by Takanaka\cite{Takanaka75} 
for uniaxial materials in the relevant order,
both except the Pauli term $w_{\rm P}$.
Thus, we have extended the results by Hohenberg and Werthamer\cite{HW67} and 
by Takanaka\cite{Takanaka75} to arbitrary crystal structures and 
impurity-scattering time, including also Pauli paramagnetism.

Equation (\ref{B_2-r_24}) reveals a close connection of both the curvature 
in $H_{c2}(T\!\alt\!T_{c})$ and the mixing of higher Landau levels in $\Delta({\bf r})$
with the Fermi-surface structure.
For example, we realize from Eq.\ (\ref{r_24}) with Eqs.\ (\ref{Wsc}), (\ref{v_+}),
and (\ref{<vFvF>}) 
that the mixing of $N\!=\! 2$ Landau level is absent for cubic materials
where $c_{1}\!=\!c_{2}\!=\! 1$.
This is not the case for low symmetry crystals, however.
Equation (\ref{B_2-r_24}) enables us to estimate 
the curvature and the mixing based on Fermi-surface structures from detailed
electronic-structure calculations.

\section{Proof of Eq.\ (\ref{P_N-BP_N})}
The first expression in Eq.\ (\ref{P_N}) can be proved by induction as follows.
First of all, $\eta_{0}\!=\!R_{0}$ is transformed from Eq.\ (\ref{R_Na}) as\cite{Wall48}
\begin{eqnarray}
&&\hspace{-5mm}
\eta_{0} = \frac{1}{1+
\displaystyle \frac{ x^{2}}{\mathstrut 1+
\displaystyle \frac{2x^{2}}{\mathstrut 1+
\displaystyle \cdots}}} 
= \frac{\sqrt{2}}{x}\,{\rm e}^{1/2x^{2}}\int_{1/\sqrt{2}x}^{\infty}{\rm e}^{-s^{2}}\, ds
\nonumber \\
&&\hspace{-0.5mm}= \int_{0}^{\infty}\exp\!\left(\!-s-\frac{x^{2}}{2}s^{2}\!\right) ds =
\frac{2}{\sqrt{\pi}}\int_{0}^{\infty}\!
\frac{{\rm e}^{-s^{2}}}{1+2x^{2}s^{2}}\,ds 
\nonumber \\
&&\hspace{-0.5mm}=
\int_{0}^{\infty}\!
\frac{2s\,{\rm e}^{-s^{2}}}{\sqrt{1+2x^{2}s^{2}}}\,ds\, .
\label{R_0}
\end{eqnarray}
Thus, Eq.\ (\ref{P_N}) holds for $N\!=\!0$.
The last expression in Eq.\ (\ref{R_0}) is the same integral 
which appears in Eq.\ (26) of Hohenberg and 
Werthamer.\cite{HW67}
We next rewrite Eq.\ (\ref{R_Na}) with respect to $\eta_{N}$ in Eq.\ (\ref{PN})
as
\begin{equation}
\eta_{N}=
\left\{ 
	\begin{array}{ll}
	\vspace{2mm}
	(-\eta_{0}+1)/x^{2} & (N= 1)
	\\
	(-\eta_{N-1}+\sqrt{N\!-\! 1}\,\eta_{N-2})/\sqrt{N}x^{2} & (N\geq 2)
	\end{array}
\right.
 \, .
\label{PN-algorithm}
\end{equation}
Using Eqs.\ (\ref{R_0}) and (\ref{PN-algorithm}), 
it is easy to see that Eq.\ (\ref{P_N}) holds for $N\!=\!1$.
Proceeding to the general case,
we assume that Eq.\ (\ref{P_N}) is valid for $N\!\leq\! M\!-\! 1$.
We also remember the following properties of the Hermite polynomials:
\begin{subequations}
\label{Hprop}
\begin{equation}
H_{N}(s)-2sH_{N-1}(s)+2(N\!-\!1)H_{N-2}(s) = 0\, , 
\label{Hprop1}
\end{equation}
\begin{equation}
\int_{0}^{\infty} s^{k}H_{N}(s)\,{\rm e}^{-s^{2}} \, ds = 0 \hspace{10mm} 
(k\!\leq \!N\!-\!1) \, .
\label{Hprop2}
\end{equation}
\end{subequations}
Then $\eta_{M}$ is obtained explicitly by using Eq.\ (\ref{PN-algorithm}) as 
\begin{eqnarray}
&&\hspace{-3mm} \eta_{M} = 
\frac{1}{\sqrt{M}x^{2}}[-\eta_{M-1}+\sqrt{M\!-\! 1}\,\eta_{M-2}]
\nonumber \\
&&\hspace{-3mm} =
\int_{0}^{\infty}\frac{2s^{M-2}[-s H_{M-1}(s)+(M\!-\!1)H_{M-2}(s)]\,{\rm e}^{-s^{2}}}{
\sqrt{\pi M!}\,x^{2}(1+2 x^{2}s^{2})}\,ds
\nonumber \\
&&\hspace{-3mm} =
\int_{0}^{\infty}\frac{-s^{M-2}H_{M}(s)\,{\rm e}^{-s^{2}}}{
\sqrt{\pi M!}\,x^{2}(1+2 x^{2}s^{2})}\,ds
\nonumber \\
&&\hspace{-3mm} =
\int_{0}^{\infty}\frac{s^{M-2}H_{M}(s)\,{\rm e}^{-s^{2}}}{
\sqrt{\pi M!}\,x^{2}}\left(1-\frac{1}{1+2 x^{2}s^{2}}\right)\,ds
\nonumber \\
&&\hspace{-3mm} =\frac{2}{\sqrt{\pi M!}}
\int_{0}^{\infty}\frac{s^{M}H_{M}(s)\,{\rm e}^{-s^{2}}}{
1+2 x^{2}s^{2}}\,ds \, .
\label{P_Nprove}
\end{eqnarray}
Thus, we have established the first expression in Eq.\ (\ref{P_N}).
The proof for the second expression proceeds in the same way by
using partial integrations
for $\eta_{0}$ and $\eta_{N-2}$ in Eq.\ (\ref{PN-algorithm}).
Equation Eq.\ (\ref{BP_N}) can be proved similarly 
by induction, starting from $\bar{R}_{1}\!=\! 1$ and using Eq.\ (\ref{R_Nb}).



\begin{thebibliography}{99}
\bibitem{comment2}More exactly, the coefficient $c_{2}$
determined by Eqs.\ (\ref{c_12-const}) 
and (\ref{chi-eq}) is chosen as real 
in the basis functions used by Rieck and Scharnberg.\cite{RS90}
\bibitem{comment3} It should be noted that $\beta$ is not independent of $\varepsilon_{n}$
but depends on the sign of $\varepsilon_{n}$. 
Thus Eq.\ (\ref{calK-sym}) may be somewhat misleading,
but it allows us a compact notation as seen from Eq.\ (\ref{Matrix}).
\end{thebibliography}
\end{document}